\documentclass[12pt]{article}
\usepackage{amssymb}
\usepackage{graphicx,color,epstopdf}
\usepackage{hyperref}
\def\hDash{\bot\!\!\!\bot}

\usepackage{amsfonts}
\usepackage{mathrsfs}
\usepackage{amsthm}
\usepackage{multirow}
\usepackage{bbding}
\usepackage{amssymb}
\usepackage{amsmath}
\usepackage{graphicx,color}
\usepackage{caption}
\usepackage{subcaption}
\usepackage{enumerate}
\usepackage{comment}
\usepackage{array}
\usepackage{bm}
\usepackage{booktabs}
\usepackage{titlesec}
\usepackage{dcolumn}

\newtheorem{theorem}{Theorem}[section]

\newtheorem{coro}{Corollary}[section]
\newtheorem{lemma}{Lemma}[section]
\newtheorem{remark}{Remark}[section]
\numberwithin{equation}{section}

\begin{document}

\title{Heteroscedasticity Testing for  Regression Models: A Dimension Reduction-based Model Adaptive Approach
\footnote{Lixing Zhu is a Chair professor of Department of Mathematics
at Hong Kong Baptist University, Hong Kong, China. He was supported by a grant from the
University Grants Council of Hong Kong, Hong Kong, China. Fei Chen was supported by a grant from the National Natural Science Foundation of China
(Grant No. 11261064).}}
\author{Xuehu Zhu$^{1}$, Fei Chen$^{2}$, Xu Guo$^{3}$ and Lixing Zhu$^1$\\
{\small {\small {\it$^1$ Hong Kong Baptist University, Hong Kong}}}\\
{\small {\small {\it$^2$ Yunnan University of Finance and Economics, Yunnan}}}\\
{\small {\small {\it $^3$ Nanjing University of Aeronautics and Astronautics, Nanjing } }}\\
}
\date{}
\maketitle

\renewcommand\baselinestretch{1.5}
{\small}

\noindent {\bf Abstract.} Heteroscedasticity testing is of importance in regression analysis. Existing local smoothing tests suffer severely from curse of dimensionality even when the number of covariates is moderate because of use of nonparametric estimation. In this paper,  a dimension reduction-based model adaptive test is proposed which behaves like a local smoothing test as if the number of covariates were equal to the number of their linear combinations in the mean regression function, in particular, equal to 1 when the mean function contains  a single index. The test statistic is asymptotically normal under the null hypothesis such that critical values are easily determined. The finite sample performances of the test are examined  by  simulations and a real data analysis.


\newpage
\baselineskip=21pt

\newpage

\setcounter{equation}{0}
\section{Introduction}
As heteroscedasticity structure would make a regression analysis more different than  that under  homoscedasticity structure, a heteroscedasticity check is required to accompany before stepping to any further analysis since ignoring the presence of heteroscedasticity may result in inaccurate inferences, say, inefficient or even inconsistent estimates.  Consider a regression model with the  nonparametric variance model:
\begin{eqnarray}\label{(1.1)}
Var(Y|X))=Var(\varepsilon |X),
\end{eqnarray}
where $Y$ is the response variable with the vector of covariates $X \in \mathbb{R}^p $ and the error term $\epsilon$ satisfies $E(\varepsilon|X)=0$.
Heteroscedasticity testing for the regression model (\ref{(1.1)}) has received much attention in the literature.
Cook and Weisberg (1983) and Tsai (1986)
proposed respectively  two score tests for a parametric structure variance function under linear regression model and first-order autoregressive model. Simonoff and Tsai (1994) further developed a modified score test under linear models. Zhu et. al. (2001) suggested a test that is based on squared residual-marked empirical process.
Liero (2003) advised a consistent test for heteroscedasticity in nonparametric regression models, which is  based on the $L^2$-distance between the underlying and hypothetical variance function. This test   is analogous to the one proposed by Dette and Munk (1998).
Dette (2002), Zheng (2009) and Zhu et. al. (2015),
extended the idea of Zheng (1996), which was primitively used for testing  mean regressions, to heteroscedasticity check under several different regression models. Further, Lin and Qu (2012) extended the idea of Dette (2002) to semi-parametric regressions. Moreover, Dette et al (2007) studied  a more general problem of testing the parametric form of the conditional variance under nonparametric regression model.



\begin{figure}
  \centering
    \includegraphics[width=\textwidth]{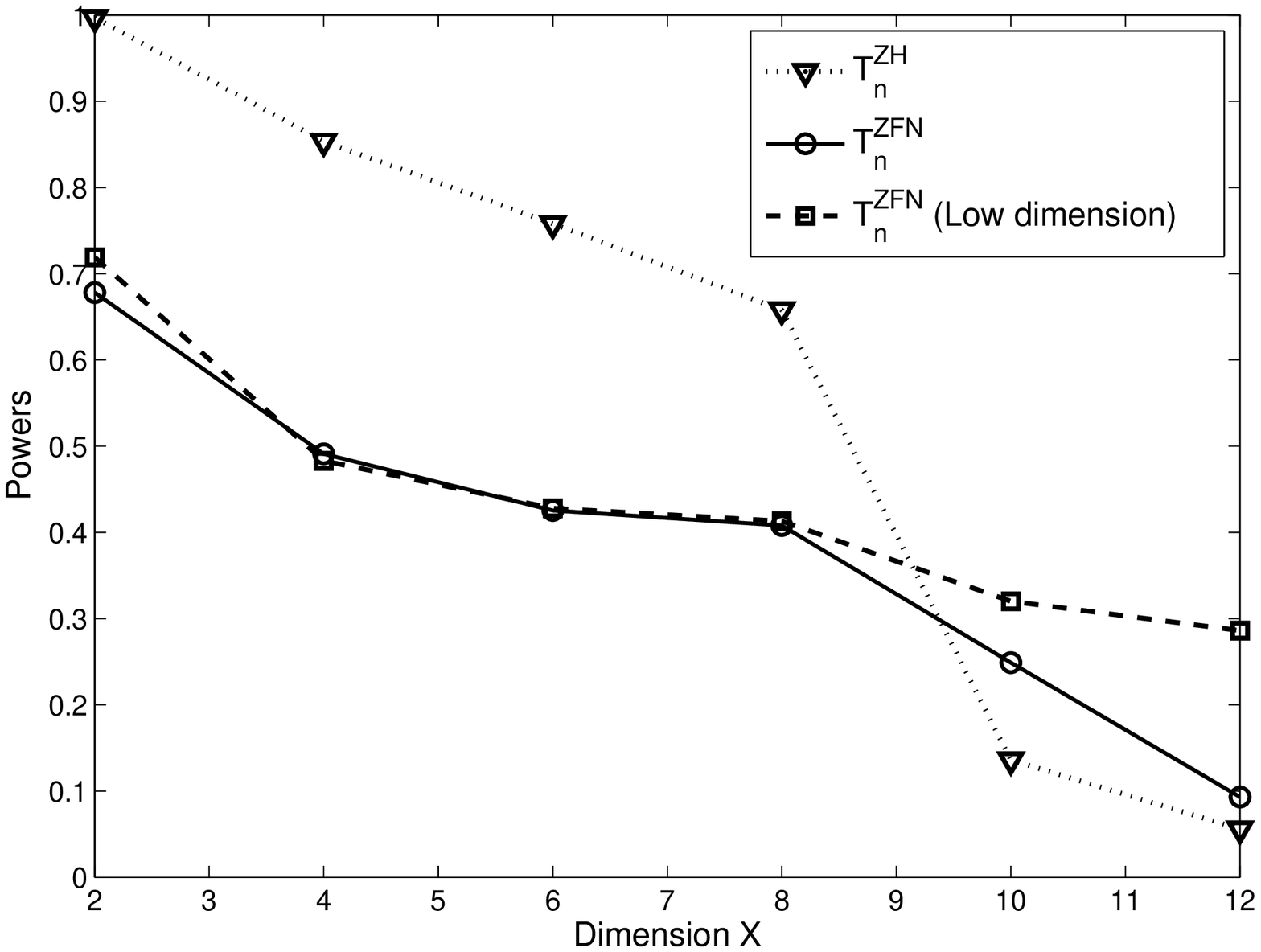}
  \caption{ The empirical power curves of Zheng (2009)'s test and Zhu et. al (2001)'s test against the dimension of $X$ with sample size 400 and $a=1$ in Example~1. Here $T^{ZFN}_n$ (low)' denotes  Zhu et. al (2001)'s test by replacing $X$ by $\beta^{\top}X$ to estimate the function $g(\cdot)$.
}\label{figure1}
\end{figure}

The  hypotheses  of interest are:
\begin{eqnarray}\label{(1.2)}
&& H_0: \exists \sigma^2>0 \mbox{ s.t. } P\{Var(\varepsilon|X) = \sigma^2\}=1 \
 \nonumber\\
 && \mbox{ against }\nonumber\\
&& H_1:  P\{Var(\varepsilon|X) =\sigma^2\}<1, \forall\
\sigma^2.
\end{eqnarray}
To motivate  the test statistic construction, we comment on  Zhu et. al. (2001)'s test and Zheng (2009)'s test as the representatives of global smoothing tests and local smoothing tests, respectively. Thanks to the fact that under the null hypothesis,
\begin{eqnarray*}
E(\varepsilon^2- \sigma^2|X)=0  \Leftrightarrow E\left\{(\varepsilon^2- \sigma^2) I(X \le t)\right\}=0 \quad \rm for\ all \quad t \in \mathbb{R}^p,
\end{eqnarray*}
 Zhu et. al. (2001) then developed a squared residual-marked empirical process as
 \begin{equation*}
V_{n}(x)= n^{-1/2}\sum_{i=1}^n\hat{\varepsilon}^2_i\{I(x_i\leq x)-F_n(x)\},
\end{equation*}
where $\hat{\varepsilon}^2_i=\{y_i- \hat{g}(x_i)\}^2$ where $\hat g (\cdot)$ is an estimate of the regression mean function. A quadratic functional form such as the Cr\"amer-von Mises type test can be constructed.
But, there exist two obvious disadvantages of this global smoothing test though it works well even when the local alternative hypotheses converge to the null hypothesis at a
rate of $O(1/\sqrt{n})$. First, it may be invalid in numerical studies of
finite samples when the dimension of $X$ is high. This is because the residual-marked empirical process for over heteroscedasticity
involves  nonparametric estimation of the mean function $g$ and thus, the curse of dimensionality severely affects the estimation efficiency.
As a local smoothing-based test, Zheng (2009)'s test can work in the scenario where the
local alternative models converge to the hypothetical model at the
rate of $O(n^{-1/2}h^{-p/4})$, where $p$ denotes the dimension of the covariate $X$ and $h$ is a bandwidth in kernel estimation. Note that the bandwidth $h$ converges to zero at a certain rate. Thus,  $O(n^{-1/2}h^{-p/4})$  can be very slow when the dimension $p$ is large.  Local smoothing tests severely suffer from the curse of dimensionality.
To illustrate those disadvantages, Figure~\ref{figure1} in Section~4 depicts the empirical powers of Zheng (2009)'s test and Zhu et. al (2001)'s test across  2000 replications with the sample size of $n=400$ against the dimension $p=2,4,6, 8,10, 12$ for a model.
This figure clearly suggests a very significant and negative impact from the dimension for the power performance of Zheng (2009)'s test and Zhu et. al (2001)'s test: when $p$ is getting larger, the power is getting down to a very low level around $0.1$ no matter the mean regression function $g(\cdot)$ is fully nonparametric or semiparametric with $\beta^{\top}X$ in the lieu of $X$. The details are presented in Section~4.

Therefore, how to handle the serious dimensionality problem  is of great importance. The goal of the present paper is to propose a new test that has a dimension reduction nature.

Consider a  general regression model in the following form:
\begin{eqnarray}\label{(1.3)}
Y=g(B^{\top}_1X)+\delta(B^{\top}_2X)e,
\end{eqnarray}
where $\varepsilon=\delta(B^{\top}_2X)e$, $B_1$ is a $p\times q_1$ matrix with $q_1$ orthonormal columns and $q_1$ is a known number satisfying $1\leq q_1 \leq p$, $B_2$ is a $p\times q_2$ matrix with $q_2$ orthonormal columns, $q_2$ is an unknown number satisfying $1\leq q_2 \leq p$, $e$ is independent of $X$ with $E(e|X)=0$  and the functions $g$ and $\delta$ are unknown.  This  model is semiparametric in the mean regression function. We assume that under the null hypothesis, the function $\delta (\cdot)$ is a constant. It is worth noting that because the functions $g$ and $\delta$ are unknown, the following  model with nonparametric variance function $\delta (\cdot)$ can also be reformulated in this form:
\begin{eqnarray*}
Y&=&g(B^{\top}_1X)+\delta(X)\varepsilon = g(B^{\top}_1X)+\delta(B_2B^{\top}_2X)e\\
&\equiv &{g} (B^{\top}_1X)+ \tilde{\delta}(B^{\top}_2X)e,
\end{eqnarray*}
where $B_2$ is any orthogonal $p\times p$ matrix. That is, $q_2=p$.  In other words, any nonparametric variance model (\ref{(1.1)}), up to the mean function, can be reformulated  as a special multi-index model with $q_2=p$. 
This model  covers many popularly used models in the literature, including the single-index models, the multi-index models and the partially linear single index models. When the model (\ref{(1.3)}) is a single index model or partially linear single index model, the corresponding number of the index becomes one or two, respectively.


In this paper, we propose a dimension reduction-based model adaptive  test (DRMAT). The basic idea is to construct a test that  is based on the local smoothing test proposed by Zheng (2009) when a model-adaptive strategy is utilized to adapt the structures under hypothetical model and alternatives. The method is motivated by Guo et. al. (2014) who considered model checks for mean regression function.  However, the construction is very different as the test not only use the model structure of conditional variance, but also the dimension reduction structure of the mean function. 
The advantages of this method include: (1) DRMAT  computes critical values by simply applying its limiting null distribution without heavy computational burden, which is often an inherent property of local smoothing testing methodologies; (2) the embedded dimension reduction procedure is model-adaptive, that is, it is automatically adaptive to the underlying model (\ref{(1.3)}) by  using more information on data such that the test can still be omnibus; more importantly, (3) under the null hypothesis,
DRMAT  has a significant faster  convergence rate of $O(n^{1/2}h^{q_1/4})$  to its limit  than $O(n^{1/2}h^{p/4})$ in existing tests when $q_1\ll p$;  and (4)
DRMAT can also detect the local alternative hypotheses converge to the null hypothesis at a much faster rate of $O(n^{-1/2}h^{-q_1/4})$ than the typical rate of $O(n^{-1/2}h^{-p/4})$. More details are presented in the next section.

 The rest of the paper is organized as follows: In Section 2, we will give a brief review for the discretization-expectation estimation and suggest a minimum ridge-type eigenvalue ratio to determine the structural dimension of the model. Moreover, the dimension reduction adaptive heteroscedasticity test is also constructed in Section 2. The asymptotic properties of the proposed test statistic under the null and alternative hypotheses are investigated in Section~3. In Section~4, the simulation results are reported and a real data analysis is carried out for illustration. Because when there are no more specific conditional variance structure assumed, the convergence rate $O(n^{-1/2}h^{-q_1/4})$ is optimal for local smoothing tests, thus, in Section~5, we discuss how to further improve the performance of DRMAT when there are some  specific conditional variance structures.  The proofs of the theoretical results are postponed to the appendix.

\section{Test statistic construction }

\subsection{Basic  construction}
Under the model~(\ref{(1.3)}), the null hypothesis is
\begin{eqnarray*}
H_0:
P\{Var(\varepsilon|X)=Var(\varepsilon|B^{\top}_1X) = \sigma^2\}=1 \quad \mbox{for some}\, \,  \sigma^2
 \end{eqnarray*}
and the alternative hypothesis
\begin{eqnarray*}
H_1: 
 P\{Var(\varepsilon|B^{\top}_2X)=
\sigma^2\}<1, \quad \mbox{for all}\, \, \sigma^2.
 \end{eqnarray*}
Write $B$ to be a $p\times q$  matrix where $q$ orthogonal columns contained in the matrix $(B_1, B_2)$.  
Then, under the null hypothesis, we have the following moment condition:
\begin{eqnarray*}
 E(\varepsilon^2-\sigma^2|X)=0.
  \end{eqnarray*}
 All existing local smoothing tests are based on this equation when the left hand side is estimated by a chosen nonparametric smoother such as kernel estimator. As was mentioned before, this severely suffers from the curse of dimensionality.  Note that, under the null hypothesis, it is unnecessary to use this technique because $E(\varepsilon^2-\sigma^2|X)=E(\varepsilon^2)-\sigma^2$. Thus, how to sufficiently use the information provided by the hypothetical model is a key to improve the efficiency of a test. It is clear that we cannot simply use two estimates in lieu of $E(\varepsilon^2)$ and $\sigma^2$ respectively to construct a test. Therefore, we consider the following idea. Note that under the null hypothesis, $B_2$ needs not to consider and thus, $B$ is reduced to $B_1$ and
\begin{eqnarray*}
 E(\varepsilon^2-\sigma^2|X)=E(\varepsilon^2-\sigma^2|B^{\top}X)=0,
  \end{eqnarray*}
and then
 \begin{eqnarray}\label{(2.1)}
E[(\varepsilon^2-\sigma^2)E(\varepsilon^2-\sigma^2|B^{\top}X)W(B^{\top}X)]=E[E^2(\varepsilon^2-\sigma^2|B^{\top}X)W(B^{\top}X)]=0,
  \end{eqnarray}
 where $W(\cdot)$ is some positive weight function which will be specified latter.
Under the alternative,
  \begin{eqnarray*}
 E(\varepsilon^2-\sigma^2|X)=E(\varepsilon^2-\sigma^2|B^{\top}X)\not=0.
  \end{eqnarray*}
  and then the left hand side of (\ref{(2.1)}) is greater than zero.
Thus, its empirical version, as a base, can be devoted to constructing a test statistic. The null hypothesis is rejected for large values of the test statistic.
We note that there are two identifiability issues.
\begin{enumerate}
\item First, for any $q\times q$ orthogonal matrix $C$ (\ref{(2.1)})holds true when the matrix $B$ is replaced by $ BC^{\top}$. This means $B$ is not identifiable while $BC^{\top}$ for an orthogonal matrix $C$ is.  But such an unidentifiabiliy problem does not affect the properties under the null and alternative hypothesis as for any $q\times q$ orthogonal matrix $C$, under the null,
\begin{eqnarray*}
 E(\varepsilon^2-\sigma^2|X)=E(\varepsilon^2-\sigma^2|CB^{\top}X)=0,
  \end{eqnarray*}
   and under the alternative hypothesis, this function is not equal to a zero function. Thus, in the following, we write $BC^{\top}$ as $B$  in the estimation procedure without notational confusion.
\item Second, under the null hypothesis, $B$ is reduced to $B_1$ with $q_1$ columns and under the alternative, $B$ has $q$ columns.
Ideally, $B$ can be identified to be $B_1$ under the null hypothesis such that we can reduce the dimension of $B$ from $q$ to $q_1$ and the nonparametric estimation can be lower-dimensional. On the other hand, under the alternative hypothesis, we wish to keep $B$ such that the criterion can fully use the information provided by the alternative and the constructed test can be omnibus. To achieve this goal, we need  a test that can automatically adapt the projected $B^{\top}X$ with the respective  dimension $q_1$ and $q$.
\end{enumerate}
 In the following estimation procedure, we introduce a model-adaptive approach.
Let $\{(x_1, y_1),\cdots (x_n, y_n)\}$ denote an $i.i.d$ sample from $(X,Y)$ and $\varepsilon_i=y_i- g(B^{\top}x_i)$.
 Then $E(\varepsilon^2-\sigma^2|B^{\top}X)$ can be estimated by the following form:
\begin{eqnarray*}
\hat{E}(\varepsilon^2-\sigma^2|\hat{B}^{\top}_{\hat{q}}x_i)=\frac {\frac{1}{(n-1)}\sum_{j\neq i,j=1}^nK_{h}\left(\hat{B}^{\top}_{\hat{q}}x_j-\hat{B}^{\top}_{\hat{q}}x_i\right)(\hat{\varepsilon}^2_j-\hat{\sigma}^2)}{\frac{1}{(n-1)}\sum_{j\neq i,j=1}^nK_{h}\left(\hat{B}^{\top}_{\hat{q}}x_j-\hat{B}^{\top}_{\hat{q}}x_i\right)}.
\end{eqnarray*}
where $\hat{\varepsilon}^2_i=(y_i - \hat{g}(\hat{B}^{\tau}x_i))^2$, $\hat{\sigma}^2=n^{-1}\sum_{i=1}^n\hat{\varepsilon}^2_i$, $K_h(\cdot)=K(\cdot/h)/h^{\hat{q}}$ with a $\hat{q}-$dimensional multivariate kernel function $K(\cdot)$, $h$ is a bandwidth and $\hat{B}_{\hat{q}}$ is an estimate of $B$ with an estimate $\hat{q}$ of $q$, which will be discussed later.
We choose the weight function $W(\cdot)$ to be the density function $p(\cdot)$ of $\hat{B}^{\top}_{\hat{q}}X$, and for any $\hat{B}^{\top}_{\hat{q}}X$, we can estimate the density function $p(\cdot)$ as the following form:
 \begin{eqnarray*}
\hat{p}(\hat{B}^{\top}_{\hat{q}}x_i)=\frac{1}{(n-1)}\sum_{j\neq i,j=1}^n K_{h}\left(\hat{B}^{\top}_{\hat{q}}x_j-\hat{B}^{\top}_{\hat{q}}x_i\right).
\end{eqnarray*}
Therefore, a non-standardized test statistic can be constructed as $S_{n}$ by:
\begin{eqnarray}\label{(2.2)}
S_{n} = \frac{1}{n(n-1)}\sum_{i=1}^n\sum_{j \neq i,j=1}^nK_{h}\left(\hat{B}^{\top}_{\hat{q}}x_i-\hat{B}^{\top}_{\hat{q}}x_j\right)(\hat{\varepsilon}^2_i-\hat{\sigma}^2)(\hat{\varepsilon}^2_j-\hat{\sigma}^2).
\end{eqnarray}
The resulting test statistic is
\begin{eqnarray}
nh^{\frac{q_1}{2}}S_{n}.
\end{eqnarray}
\begin{remark}\label{Remark2.1} From the construction, it seems that except an estimate of $B$, the test statistic has no difference in spirit from that  by Zheng (2009) as follows:
\begin{eqnarray}\label{(2.3)}
\tilde{S}_{n} = \frac{1}{n(n-1)}\sum_{i=1}^n\sum_{j \neq i,j=1}^n\tilde{K}_{h}\left(x_i-x_j\right)(\hat{\varepsilon}^2_i-\hat{\sigma}^2)(\hat{\varepsilon}^2_j-\hat{\sigma}^2),
\end{eqnarray}
where $\tilde{K}_{h}(\cdot) = \tilde{K}(\cdot/h)/h^{p}$ with a $p$-dimensional multivariate kernel function $\tilde{K}(\cdot)$. Our test statistic is even more complicated in the case where $q=p$  because even $q$ is given, our test still involves an estimate of $q$. However, we note that if we do not have this estimate that can automatically adapt to $q_1$ and $q$, we do not have chance to construct a test that can use the standardizing constant $nh^{\frac{q_1}{2}}$.  Zheng (2009)'s test statistic must use, actually all existing local smoothing tests, the standardizing constant  $nh^{\frac{p}{2}}$, otherwise, the limit goes to infinity if $nh^{\frac{q_1}{2}}$ is used. Comparing  (\ref{(2.2)}) with (\ref{(2.3)}),  we observe that the different dimensions of the kernel estimators in $S_{n}$ and $\tilde S_{n}$ make this significant improvement of $S_{n}$ than  $\tilde S_{n}$. Clearly, under the null hypothesis, the curse of dimensionality is largely avoided. As we will see in Section~3, $S_n$ is asymptotically normal at the rate of order $nh^{q_1/2}$ under the null hypothesis, whereas $\tilde S_{n}$ has the asymptotic normality at the rate of order $nh^{p/2}$. Particularly, when the model (\ref{(1.3)}) is a single-index model or generalized linear model, $q_1=1$.  More importantly, in Section~3, we will show that our test can be much more sensitive than existing local smoothing tests in the sense that it can detect local alternatives converging to the null at the rate of $1/(\sqrt{ n}h^{q_1/4})$ that is a much faster rate than $1/(\sqrt{n}h^{p/4})$ existing local smoothing tests can achieve.  This gain is again due to the adaptive estimation of  the matrix $B$ under the null and alternative respectively.
In the next section, we will use sufficient dimension reduction technique to construct an estimate of $q$.  
\end{remark}

\begin{remark}\label{Remark2.1}
In the construction, we consider the conditional expectation given  $B^{\top}X$ where $B$ consists of both $B_1$ and $B_2$. Under the null, $B$ is reduced to $B_1$, but under the alternative the dimension $q$ can be greater than $q_2$ and thus, we have to use higher order kernel in local smoothing step.  A natural idea is to estimate $B_1$ and $B_2$ separately. However, we find that such a procedure makes the implementation more complicated. Thus, in the present paper, we do not use this idea and a further research is ongoing to see how to improve the performance of the test.
\end{remark}

\subsection{A review on discretization-expectation estimation}

As illustrated in Subsection 2.1, estimating $B$ plays an important role for the  test statistic construction. To this end, we consider sufficient dimension reduction technique.  Since $B$ is not identifiable in the model~(\ref{(1.3)}) and the
functions $g(\cdot)$ and $\delta(\cdot)$ are unknown, what we can identify is $BC^{\top}$ because for
any $q \times q$ orthogonal matrix $C$, $g(B^{\top}X)$ and
$\delta(B^{\top}X)$ can be further rewritten as
$\tilde{g}(C^{\top}B^{\top}X)$ and
$\tilde{\delta}(C^{\top}B^{\top}X)$. Sufficient dimension reduction technique helps us identify the
subspace spanned by $B$ called  the central subspace (Cook 1998).  More precisely,
from the definition of the central subspace, it is the intersection of all subspaces $\emph{S}$ such that
$$Y \hDash X|(P_{\emph{S}}X),$$
where $\hDash$ denotes the statistics independence and $P_{(\cdot)}$ stands for a projection operator with respect to the standard inner product. $dim(P_{\emph{S}}X)$ is called the structure dimension of $P_{\emph{S}}X$ and is $q$ in our setup. In other words, $P_{\emph{S}}$ is equal to $CB^{\top}$ for some $q\times q$ orthogonal matrix $C$. As we mentioned, we still use $B$ without confusion.

There exist several promising dimension reduction proposals available in the literature. For example, Li (1991) proposed  sliced inverse regression (SIR), Cook and Weisberg (1991) advised sliced average variance estimation (SAVE), Xia et al, (2002) discussed minimum average variance estimation (MAVE),
and Zhu et al. (2010) suggested discretization-expectation estimation (DEE).
As  DEE does not need to select the number of slices and has been proved in the simulation studies to have  a good performance,  we then adopt it to estimate $B$ for the test statistic construction. From Zhu et al (2010), the SIR-based DEE can be carried out by the
following  estimation steps.
\begin{enumerate}
\item Discretize the response variable $Y$ into a set of binary variables by defining
 $Z(t)= I\{Y \leq t\}$, where the indicator function $I\{Y \leq t\}$ takes  value 1 if $I\{Y \leq t\}$ and 0 otherwise.
\item Let $\emph{S}_{Z(t)|X}$ denote the central subspace of $Z(t)|X$. When SIR is used, the related SIR matrix  $M(t)$ is  an $ p\times  p$ positive semi-definite matrix satisfied that $\rm{Span}\{M(t)\} =\emph{S}_{Z(t)|X}$.
\item Let $\tilde Y$ be an independent copy of $Y$. The target matrix is $M=E\{M(\tilde Y)\}$. $ B$ consists of the eigenvectors associated with the nonzero eigenvalues of $M$.
\item Obtain an estimate of $M$ as:
\begin{equation*}
M_{n}=\frac{1}{n}\sum^{n}_{i=1}M_n({y}_i),
\end{equation*}
where $M_n({y}_i)$ is the estimate of the SIR matrix $M({y}_i)$. When $q$ is given,  an estimate $ \hat B_q$ of $B$ consists of the eigenvectors  associated with the largest $q$ eigenvalues of $M_{n}$.  $ \hat B_q$ can be root-$n$
consistent to $B$. More details can be referred to Zhu et. al. (2010).
\end{enumerate}

\subsection{The estimation of structural dimension}
To completely estimate $B$, we also need to estimate the structural dimension $q$.  Thus, an estimate of $q$ is essential for the test statistic. The BIC-type criterion was suggested by Zhu et. al. (2010). However,  choosing an appropriate tuning parameter is an issue.  Thus, we suggest another method that is very easy to implement. In the aforementioned sufficient dimension reduction procedure, the estimating matrix $M_{n}$ is a root-$n$ consistent estimation of the target matrix $M$. Let $\hat{\lambda}_{p} \leq \hat{\lambda}_{(p-1)}\leq \cdots \leq \hat{\lambda}_{1}$ be the eigenvalues of the estimating matrix $M_n$.
In spirit similar to that in Xia et al (2014),
we advise a  ridge-type eigenvalue ratio estimate (RERE) to determine $q$ as:
\begin{eqnarray}\label{(2.4)}
\hat{q}=\arg\min_{1\leq j \leq p}\left\{ \frac{\hat{\lambda}^2_{j+1}+c_n}{\hat{\lambda}^2_j+c_n}\right\}.
\end{eqnarray}
The following theorem shows that the structure dimension $q$ can be consistently determined by RERE criterion.
\begin{theorem}\label{theorem0}
Under Conditions A1 and A2 in Appendix,  the estimate $\hat{q}$ of (\ref{(2.4)}) with $\frac{\log{n}}{n} \leq c_n \rightarrow 0$ satisfies that as $n \rightarrow 0$ in a probability going to $1$,
\begin{itemize}
\item [(i)] under $H_0$, $\hat{q} \rightarrow q_1$;
\item [(ii)] under $H_1$, $\hat{q} \rightarrow q$.
\end{itemize}
\end{theorem}
This theorem implies that, in the test statistic construction, the structure dimension estimate $\hat{q}$ can be automatically adaptive to the  model (\ref{(1.3)}) rather than the nonparametric model (\ref{(1.1)}).  An consistent estimate of $B$ is denoted by $\hat{B}_{\hat{q}}$. In the above test statistic construction, this estimate plays a crucial role as $\hat{B}_{\hat{q}}$ converges to a $p\times q$ matrix $B$ $B$ under $H_1$  and to a $p\times q_1$ matrix $B$ under $H_0$.

\section{Asymptotic properties}
\subsection{Limiting null distribution}
Define two notations first. Let
\begin{eqnarray}\label{(3.1)}
s^2 = 2\int{}K^2(u)duE\{[Var(\varepsilon^2|B^{\top}X)]^2p(B^{\top}X)\},
\end{eqnarray}
and
\begin{eqnarray}\label{(3.2)}
\hat{s}^2 = \frac{2}{n(n-1)}\sum_{i=1}^n\sum_{j \neq i,j=1}^nK^2_h\left(\hat{B}^{\top}_{\hat{q}}x_i-\hat{B}^{\top}_{\hat{q}}x_j\right)(\hat{\varepsilon}^2_i-\hat{\sigma}^2)^2(\hat{\varepsilon}^2_j-\hat{\sigma}^2)^2.
\end{eqnarray}
We will prove that $\hat{s}^2$  is a consistent estimate of  $s^2$ under the null and local alternative hypotheses. Further, we have the following asymptotic properties of the test statistic under the null hypothesis.

\begin{theorem} \label{theo3.1} Given Conditions A1-A8 in Appendix and under $H_0$, we have
$$nh^{\frac{q_1}{2}}S_{n} \stackrel{\mathrm{d}}{\rightarrow} N(0,s^2),$$
where the notation $ \stackrel{\mathrm{d}}{\rightarrow}$ denotes convergence in distribution and $s^2$ is defined by (\ref{(3.1)}).
Further, $s^2$ can be consistently estimated by $\hat{s}^2$ given by (\ref{(3.2)}).
\end{theorem}


According to Theorem~\ref{theo3.1}, by standardizing $S_{n}$, we then get an standardized test statistic $T_{n}$ as:
\begin{eqnarray}\label{(3.3)}
T_{n} = nh^{{q_1}/2}S_{n}/\hat{s}.
\end{eqnarray}
 Furthermore, using the Slusky theorem yields the following corollary.
\begin{coro}\label{coro1}
Under the conditions in Theorem \ref{theo3.1} and $H_0$, we have
\begin{eqnarray*}
T_{n} \stackrel{\mathrm{d}}{\rightarrow} N(0,1),
\end{eqnarray*}
and then
\begin{eqnarray*}
T^2_n \stackrel{\mathrm{d}}{\rightarrow} \chi^2_1,
\end{eqnarray*}
where $\chi^2_1$ is the chi-square distribution with one degree of freedom.
\end{coro}
Based on Theorem \ref{theo3.1} and Corollary~\ref{coro1}, it is easy to calculate $p-$values by its limiting null distribution of $T_n^2$. As a popularly used approach, Monte Carlo simulation can also be employed.

Summarizing all the aforementioned constructing procedure gives the following steps:
\begin{itemize}
\item[Step] 1: Use a sufficient dimension reduction method such as DEE
(Zhu et. al. 2010) to detain the estimators $\hat{B}_{\hat{q}}$ with $\hat{q}$ determined by MRRE criterion, and then apply the nonparametric kernel method to estimate the mean function $g(\cdot)$ as follows:
\begin{eqnarray*}
\hat{g}(\hat{B}^{\top}_{\hat{q}}x_i)=\frac{\sum_{i=1}^nQ_{h_1}\left(\hat{B}^{\top}_{\hat{q}}x_j-\hat{B}^{\top}_{\hat{q}}x_i\right)y_i}
{\sum_{i=1}^nQ_{h_1}\left(\hat{B}^{\top}_{\hat{q}}x_j-\hat{B}^{\top}_{\hat{q}}x_i\right)},
\end{eqnarray*}
where $Q_{h_1}(\cdot)=Q(\cdot/h_1)/h_1$ with $Q(\cdot)$  being a $\hat{q}-$dimensional kernel function and $h_1$ being a
bandwidth.
\item[Step] 2: Calculate the test statistic $T_{n} = nh^{q_1/2}S_{n}/\hat{s}$ with $S_n$ and $\hat{s}$ given by (\ref{(2.2)}) and (\ref{(3.2)}), respectively.
\end{itemize}

\subsection{Power study}
To  study the power performance of the proposed test statistic, consider the sequence of local alternative hypotheses with the following form:
\begin{eqnarray}\label{(3.4)}
H_{1n} : Y=g(B^{\top}_1X)+\eta, Var(\eta|X)= \sigma^2 +C_nf(B^{\top}X)
\end{eqnarray}
where $f(\cdot)$ is some continuously differentiable function  satisfying $E[f^2(X)] < \infty$ and the columns of $B_1$ can be the linear combination of the columns of $B$.

Under the global alternative with fixed $C_n$, Theorem~\ref{theorem0} shows that  $\hat q$ converges to $q$ in a probability going to $1$. Under the local alternative, $C_n$ goes to zero and the part with $B_2$ vanishes as $n\to \infty$, we may expect that $\hat q$ also tends to $q_1$ although at the population level, the structural dimension is still $q$. In other words, the estimate $\hat q$ is not consistent. But this is just what we want because it will make the test more sensitive to the local alternative. The following states this result.

\begin{lemma}\label{lemma1} Under the local alternative $H_{1n}$ in $(\ref{(3.4)})$ with $C_n = n^{-\frac{1}{2}}h^{-\frac{q_1}{4}}$ and the same conditions in Theorem~\ref{theorem0} except that $C_n^2\log{n} \leq c_n \rightarrow 0$,  the estimate $\hat{q}$ given by (\ref{(2.4)}) satisfies that $\hat{q}\rightarrow q_1$  in probability as $ n\rightarrow 0 $.
\end{lemma}

Now, we state  the power performance of the test.
\begin{theorem} \label{theo2.2} Under Conditions A1-A8 in Appendix, we have the following results.
\begin{itemize}
\item[(I)] Under the global alternative hypothesis $H_1$,  we have
$$ S_{n} {\stackrel{\mathrm{p}}{\rightarrow}} E\big{\{}[Var(\varepsilon|B^{\top}X)-Var(\varepsilon)]^2p(B^{\top}X)\big{\}},$$
and
$$\hat s^2 \stackrel{\mathrm{p}}{\rightarrow} 2\int{}K^2(u)duE\big{\{}[Var(\varepsilon^2|B^{\top}X)+(Var(\varepsilon|B^{\top}X)-\sigma^2)^2]^2p(B^{\top}X)\big{\}},$$
where the notation $\stackrel{\mathrm{p}}{\rightarrow}$ denotes convergence in probability and $\hat{s}^2$ is defined in (\ref{(3.2)}). Thus,
 $$ T_{n}/(nh^{q_1/2}) \stackrel{\mathrm{p}}{\rightarrow} Constant.$$
\item[(II)] Under the local alternative hypothesis $H_{1n}$ in (\ref{(3.4)}) with $C_n = n^{-\frac{1}{2}}h^{-\frac{q_1}{4}}$, we have
$$ T_{n} \stackrel{\mathrm{d}}{\rightarrow} N(m,1),$$
and
$$ T^2_{n} \stackrel{\mathrm{d}}{\rightarrow} \chi^2_1(m^2),$$
where $ m= E\{[E\{f(B^{\top}X)|B^{\top}_1X\}]^2p(B^{\top}_1X)\}/s $ with $s$ given by (\ref{(3.1)}) and $ \chi^2_1(m^2)$ is a noncentral chi-squared random variable with one degree of freedom and the noncentrality parameter $m$.
\end{itemize}
\end{theorem}


\begin{remark}\label{Remark2.}
The above results confirm the claims we made in Section~1. Unlike the convergence rates in Zheng (2009), our test can have the following rates under the null and alternative hypothesis. (I) Under the null hypothesis, $S_n$ converges to its limit at the rate of order $nh^{q_1/2}$ whereas $\tilde{S}_{n}$ has a much slower rate of order $nh^{p/2}$. Particularly, when the null hypothesis belongs to the single index models or the generalized linear models, $S_n$ has a fastest convergence rate as if the dimension $p=1$, namely, $nh^{1/2}$. (II)  $T_n$ can detect the local alternative models converging to the hypothetical model at the rate of order $n^{-\frac{1}{2}}h^{-\frac{q_1}{4}}_1 $ rather than $n^{-\frac{1}{2}}h^{-\frac{p}{4}}_1 $ that Zheng's test can achieve. 
\end{remark}

\section{Numerical Studies}
\subsection{Simulations}
In this subsection, we conduct the following simulations to illustrate the performance of the proposed test. We choose the product of $\hat{q}$ Quartic kernel function as $K(u) = Q(u)=15/16(1 - u^2)^2$, if $|u| \leq 1$ and 0 otherwise. In the test statistic, $B$ is estimated by  the SIR-based DEE procedure  and $q$ is used by the RERE criterion (\ref{(2.4)}) with $c_n=\log{n}/(nh^{q_1/2})$.  Let $T^{DEE}_{n}$,
$T^{ZH}_{n}$ and $T^{ZFN}_{n}$ denote  the proposed test in the present paper,
Zhu et. al. (2002)'s test and Zheng (2009)'s test, respectively. We focus on the
performance of these tests under different settings of the dimension
and the correlation structure of the covariate vector $X$ and the
distribution of the error term $\varepsilon$. The sample sizes are $50, 200, 400.$ The
empirical sizes and powers are computed through 2000 replications for each experiment
at the significance level $\alpha=0.05$.


The observations $x_i$, for $i=1,2, \cdots, n$  are i.i.d. from multivariate normal distribution $N(0, \Sigma_1)$ or $N(0, \Sigma_2)$ and independent of
the standard normal errors $\varepsilon$, where
$\Sigma_1=(\sigma_{ij}^{(1)})_{p\times p}$ and
$\Sigma_2=(\sigma_{ij}^{(2)})_{p\times p}$ with the elements respectively
\begin{eqnarray*}
\sigma_{ij}^{(1)}=I(i=j)+ 0.5^{|i-j|} I(i \neq j) \ \
{\rm and} \ \ \sigma_{ij}^{(2)}=I(i=j)+ 0.3I(i \neq j).
\end{eqnarray*}

\begin{figure}
  \centering
  \includegraphics[width=\textwidth]{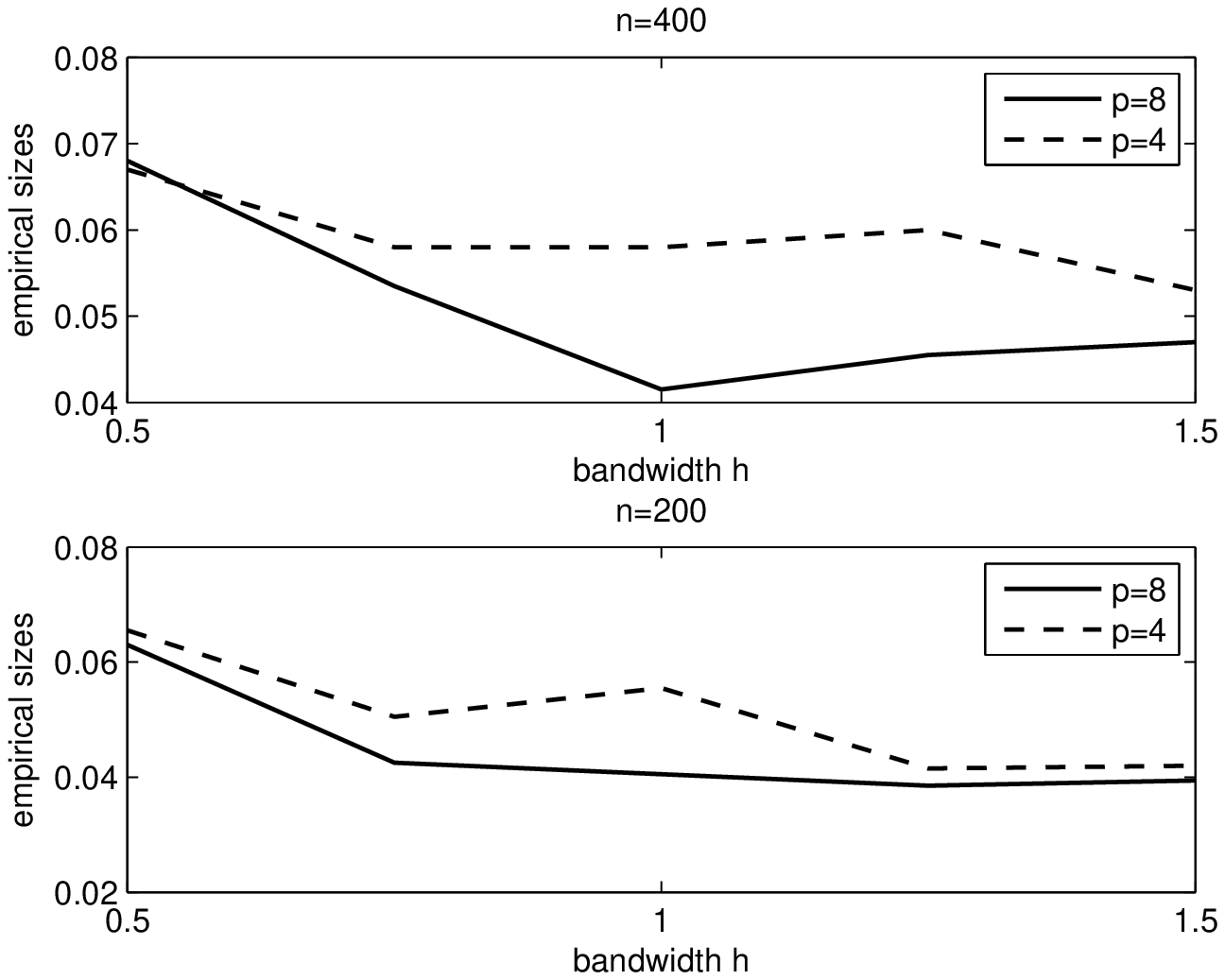}\\
  \caption{The empirical size curves of $T^{DEE}_n$ against the bandwidth  and sample size 200 (the below panel) and 400 (the above panel) with $a=0$ in Example~1}\label{figure2}
\end{figure}

\begin{figure}
  \centering
  \includegraphics[width=\textwidth]{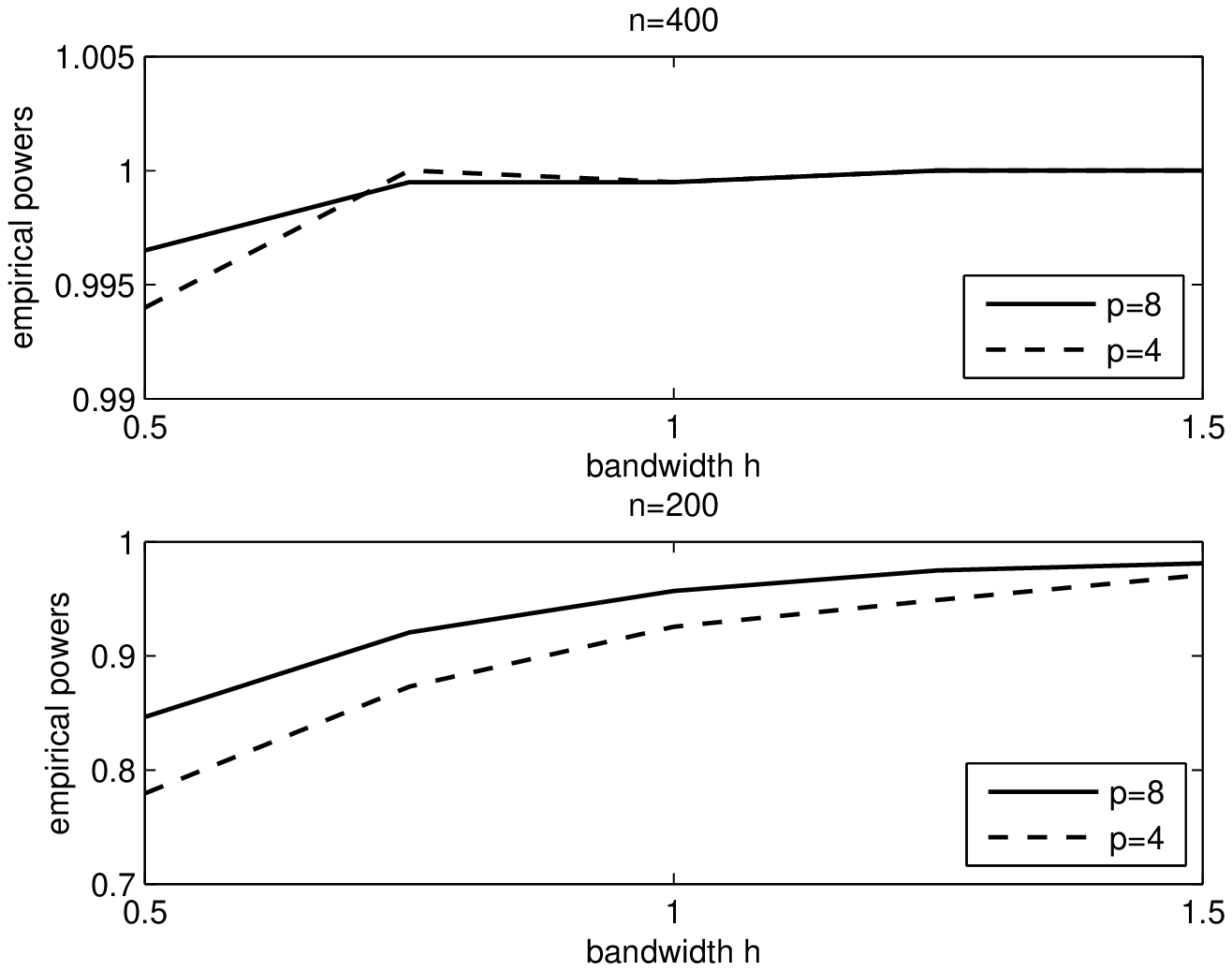}\\
  \caption{The empirical power curves of $T^{DEE}_n$ against the bandwidth  and sample size 200 (the below panel) and 400 (the above panel) with $a=1$ in Example~1}\label{figure3}
\end{figure}

\textbf{Example 1.} Consider the following single-index model:
 \begin{eqnarray*}
Y=\beta^{\top}X +\exp(-(\beta^{\top}X)^{2})+ 0.5(1 + a \times |\beta^{\top}X|)\times\epsilon,
 \end{eqnarray*}
where  $X$  follows normal distribution $N(0, \Sigma_1)$, independent of the standard normal errors $\epsilon$ and $\beta=(\underbrace{1,\cdots,1}_{p/2},0,\cdots,0)^\tau/\sqrt{p/2}$ and $p$ is set to be 2, 4 and 8 to  reveal the impact from dimension. Moreover,  $a =0$ and $a \neq 0$  respectively, correspond to the null and the alternative hypothesis. First, we investigate the impact from bandwidth selection.
We choose different bandwidths $(0.5 + 0.25\times i)n^{-1/(4+\hat{q})}$ for $i = 0, \cdots, 5$ and obtain the empirical sizes and powers when the dimension of $X$ is relatively high, namely, $p=4, 8$.
\begin{center}
Figures \ref{figure2} and \ref{figure3} about here
\end{center}

From Figures  \ref{figure2} and \ref{figure3},  it can be clearly observed that the test is  robust against different bandwidths and the type I error can be controlled well. On the other hand, when  the sample size is small,  the bandwidth selection has little impact for the power performance when the bandwidth becomes small.  Therefore, we recommend $h=1.25\times n^{-1/(4+\hat{q})}$ in the simulations.

Now we turn to compare the empirical sizes (type I errors) and powers of all the three tests under different combinations of sample sizes and dimensions of covariate vector $X$.
The results are reported in Table \ref{table1}.
\begin{center}
Table \ref{table1} about here
\end{center}
It is clear that the empirical power increases as $a$ gets larger. Further,  our test $T^{DEE}_n$
is significantly more powerful than  $T^{ZH}_{n}$ and $T^{ZFN}_{n}$. In addition, $T^{DEE}_n$ can control the size very well for the different dimensions of $X$ and the sample sizes. However, for $T^{ZH}_{n}$ and $T^{ZFN}_{n}$, the impact from the dimension of $X$ is very significant. When the dimension of $X$ becomes larger, the empirical
power of our test  slightly changes whereas the empirical
powers of $T^{ZH}_{n}$ and $T^{ZFN}_{n}$ drop down quickly. Even when the sample size is $n=400$, the situation does not become significantly better.  This implies that the
dimensionality is a big obstacle for these two tests to perform well.

Next,  we use the following example that has two different $B$ under the null and alternative hypothesis to check the usefulness of model-adaptiveness in promoting the performance of test. Further, we examine the robustness of the test against different error $\varepsilon$.

\textbf{Example 2.} Consider the following model in which the dimension of $B_1$ and that of $B$ are different under $H_1$:
 \begin{itemize}
 \item  $Y= \beta^{\top}_1X +  0.5 [a\times\{(\beta^{\top}_1X)^2+(\beta^{\top}_2X)^2\}+1]\times \epsilon $.
 \end{itemize}
where  $\beta_1=(1,1, 0,0)^\tau/\sqrt{2}$ and $\beta_2=(0,0,1,1)^\tau/\sqrt{2}$. Further, $\epsilon$ follows the Student's t-distribution $t(6)$ with degrees of freedom 6. Two cases are investigated, where  $X$ follows
$N(0,\Sigma_1)$ and $N(0,\Sigma_2)$, respectively. In this example, under $H_0$ with $a=0$, $B=\beta_1$ and under $H_1$ with $a\not = 0$, $B=(\beta_1, \beta_2).$
The  results are presented in Table \ref{table2}.
\begin{center}
Table \ref{table2} about here
\end{center}
From Table~\ref{table2}, we have the following findings. First, the comparison between the two cases of this example shows that the correlation structure of  $X$ would not deteriorate the power performance. Second, in the limited simulations, the heavy tail of the error term does not have a significant impact on the performance of our test. Third, we can observe that although $B$ has higher dimension $q=2$ under  the alternative hypothesis than $q_1=1$ under the null in
this example, the results are still similar to those in Example~1 that has the same $B$ in the hypothetical and alternative model.
These findings suggest that the proposed test is robust against the correlation structure of $X$ and the different error $\varepsilon$. The power performance is less negatively affected by the dimension under the alternative model.

\textbf{Example 3.} To further examine the performance of te proposed test, consider the following model:
\begin{itemize}
 \item  $Y= \beta^{\top}_1X + 2\sin(\beta^{\top}_2X/2) + 0.5 [a \times\{(\beta^{\top}_1X)^2+(\beta^{\top}_2X)^2\}+1]^{0.5}\times \epsilon $.
 \end{itemize}
where the values of  $\beta_1$, $\beta_2$ and $p$ are set to be the same as those in example 2  and $\epsilon$ is from normal distribution. In this example, $B$ is identical under the null and alternative hypotheses as that in Example~1, but $q=q_1=2.$
\begin{center}
Table \ref{table3} about here
\end{center}
Because the results of $T^{ZH}_{n}$ and $T^{ZFN}_{n}$ are similar to those in Example~2, we omit to present the detailed results and only present the results of our test $T^{DEE}_{n}$ to save the space.  By the comparison between Tables~2 and 3,
we can find that the test power with $q_1=2$ in Example~3  is lower than that with $q_1=1$ in Example~2.

These numerical results support the aforementioned
theoretical results indicating that DRMAT has significantly improved the performance of
existing local smoothing tests.  The empirical sizes also
show that, in our test, critical values computed by
simply applying the limiting null distribution is reliable. Hence, the
computational workload of DRMAT  is not heavy.

\subsection{Real Data Analysis}
We consider the well-known 1984 Olympic records data on various track events, which has been analyzed by Naik and Khattree (1996) using the method of principal component analysis  for the investigation of their athletic excellence and the relative strength on certain countries at the different running. Further, Zhu (2003) once analyzed
this data in checking certain parametric structure. The data for men consists of 55 countries with eight running events presented, which are the 100 meters, 200 meters, 400 meters, 800 meters, 1,500 meters, 5,000 meters, 10,000 meters and the Marathon distance, see Naik and Khattree (1996).

\begin{figure}
  \centering
    \includegraphics[width=7cm]{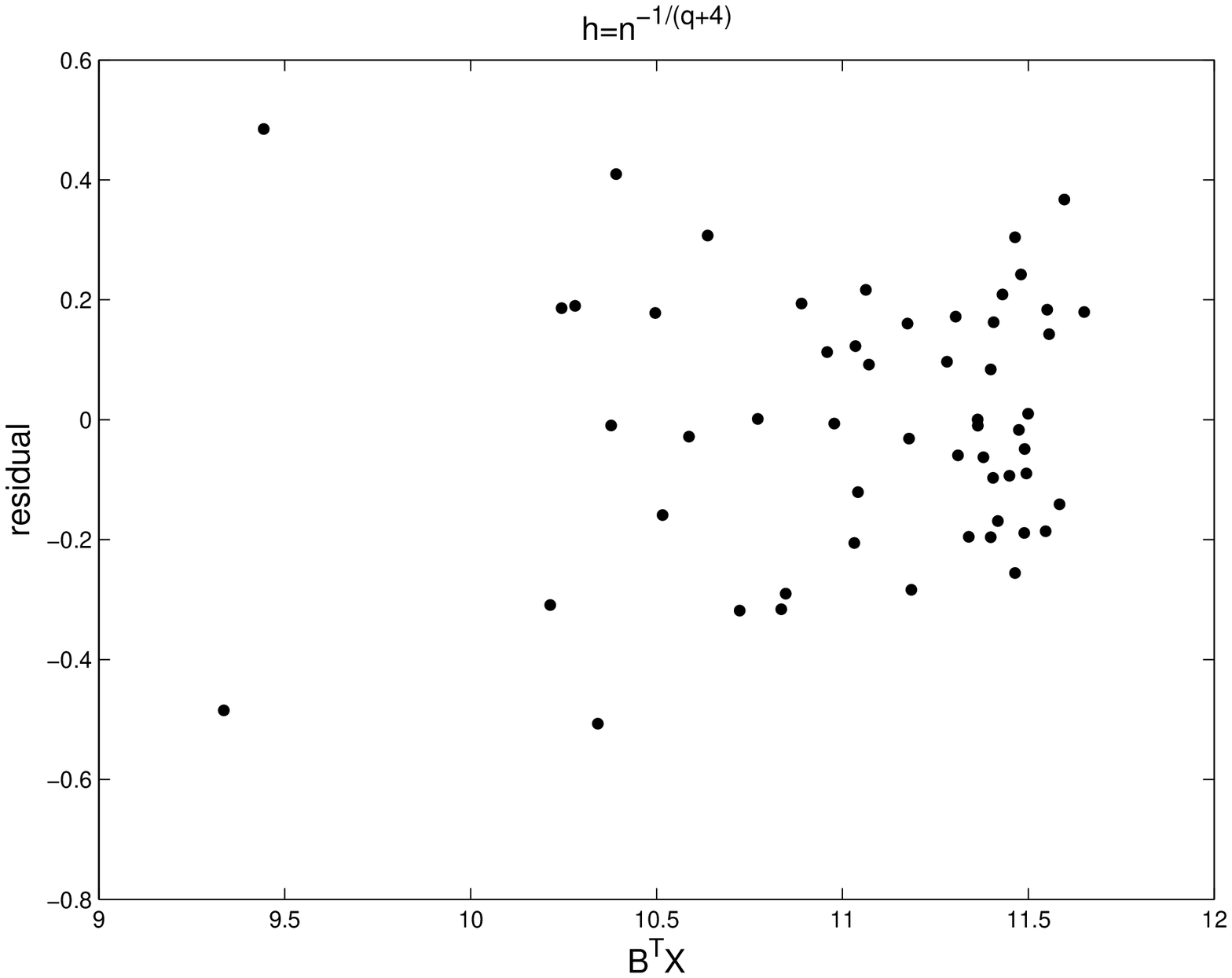}
  \includegraphics[width=7cm]{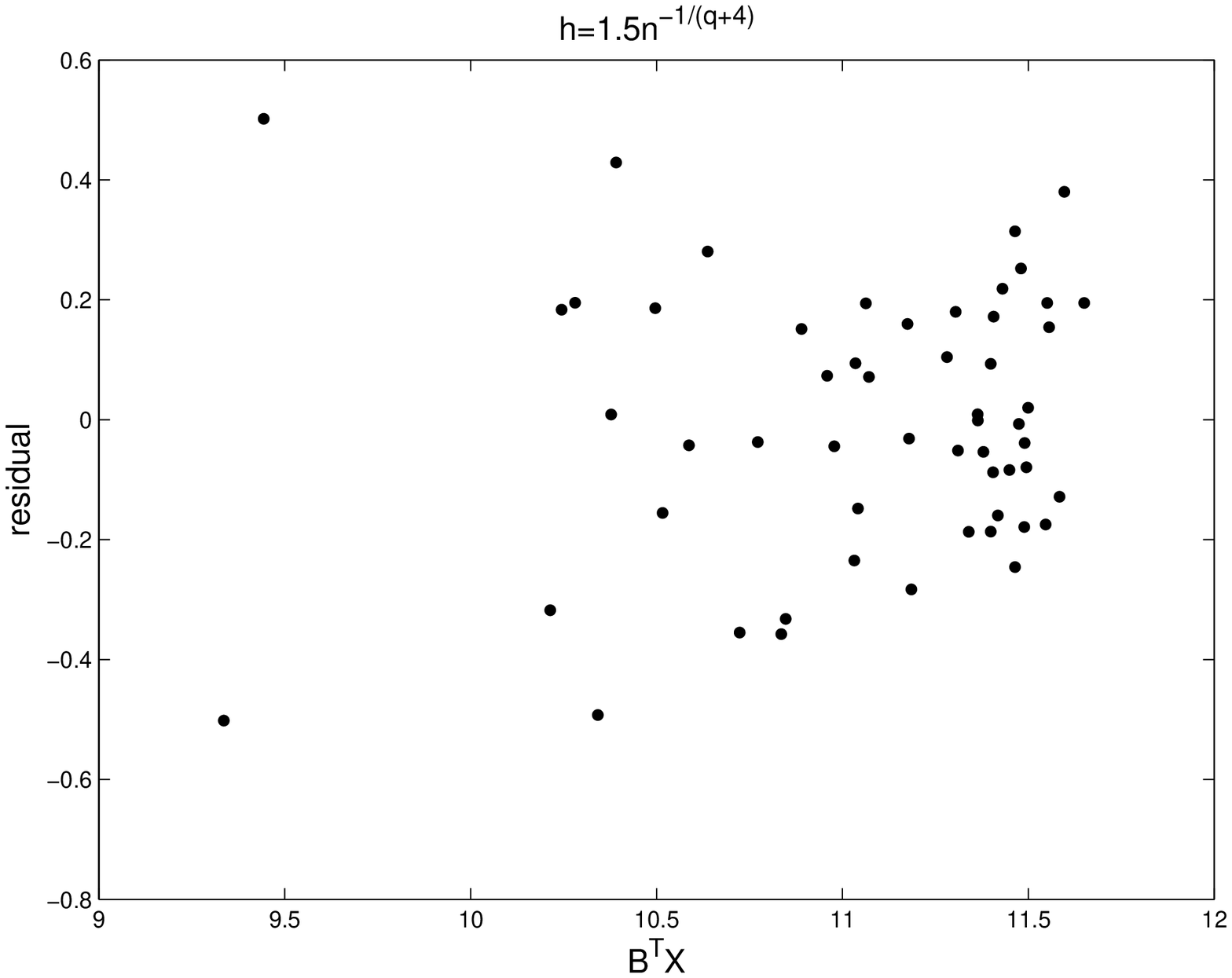}
  \caption{ The residual plots from the regression model (\ref{(1.3)}) against the
single-indexing direction obtained from DEE in the real data analysis.
}\label{figure4}
\end{figure}

As argued by Naik and Khattree (1996), it may be more tenable to use the speed rather than the winning time for the study. Here, what we are interested in is to examine whether the performance of a nation in running long distances  has a significant  effect on  that in short running speed, see Zhu (2003). We also take the speed of the 100 meters running event as the response and the speed of the 1,500 meters, 5,000 meters, 10,000 meters and the Marathon distance as covariates.
\begin{center}
Figure~\ref{figure4} about here
\end{center}
Figure (\ref{figure4}) presents the plots of the residuals versus $B^{\top}X$ with the different  bandwidth $h=n^{-1/(4+\hat{q})}$ and $h=1.5n^{-1/(4+\hat{q})}$, where the estimator of $ B$ is constructed by DEE and $q$  by RERE.
From the plots  we  see that a heteroscedasticity structure may exist.
The values of the test statistic are computed as $T_1 =  3.4230$ and $T_2 = 3.9028$ with $h=n^{-1/(4+\hat{q})}$ and $h=1.5n^{-1/(4+\hat{q})}$, respectively and the corresponding $p-$values are $0.0003$ and $0.0000$. Thus, a heteroscedasticity model may be tenable for this data set.
The result of this analysis implies that the volatility of the performance in running short distances depends on the performance in running long distances.

\section{Discussions}
Heteroscedasticity checking is an important step in regression
analysis. In this paper, we develop a dimension
reduction model-adaptive  test. The critical ingredient in the test statistic construction is that the test embeds the dimension reduction
structure under the null hypothesis to overcome the curse of dimensionality and adopts to model structure under the alternative such that it
 is still an omnibus test. The test statistic has the limit at the rate as if the number of covariates was the number of linear combinations in the mean regression  function. Note that under the null hypothesis, the number of covariates is $0$. Thus could we further improve our test to have a faster rate? Looking at the construction procedure, it seems not possible if we do not have any other extra assumptions on the conditional variance structure. However, if we have  prior information that under the alternative hypothesis, an improvement seems possible. For instance, when we know that $q$ is greater than $q_1$, the consistency of the estimator $\hat q$ gives us the chance to have idea whether the underlying model is hypothetical or alternative model. Of course, we cannot simply use this information to be a test as type I and II errors cannot be determined. We then use an estimate $\hat B_{\tilde q}$ where $\tilde q=I(\hat q=q_1)+\hat qI(\hat q>q_1)$ and the test statistic is based on $\hat B_{\tilde q}$. This means that under the null hypothesis, with a probability going to $1$, the test statistic is only with one linear combination of the covariates, rather than $q_1$ linear combinations. The standardizing constant is $nh^{1/2}$ rather than $nh^{q_1/2}$. It is expectable to have the asymptotic normality under the null hypothesis. It can also detect the local alternatives distinct from the null at the rate of $1/\sqrt {nh^{1/2}}$. The study is ongoing.

 Further, this method can also be extended to handle other conditional variance models such as
single-index and multi-index models.  The relevant research is ongoing.

\section{Appendix.}

\subsection{Regularity Conditions}
To investigate the theorems in Section 3, the following regularity conditions are designed.
\begin{itemize}
\item [A1] $M_n(t)$ has the following expansion:
    $$M_n(t)=M(t)+E_n\{\psi
    (X,Y,t)\}+R_n(t),$$where $E_n(\cdot)$ denotes sample averages,
    $E(\psi (X,Y,t))=0$ and $\psi(X,Y,t)$ has a finite second-order moment.
\item [A2]$\sup_{t\in R^{p_2}} ||R_n(t)||_F=o_p(n^{-1/2})$, where $||\cdot||_F$ denotes the
    Frobenius norm of a matrix.
\item[A3] $(B^{\top}x_i,y_i)_{i=1}^n$ follows a probability distribution $F(B^{\top}x,y)$ on $\mathbb{R}^q\times \mathbb{R}$. $E(\varepsilon^8|B^{\top}X=B^{\top}x)$ is continuously differentiable and $E(\varepsilon^8|B^{\top}X=B^{\top}x)\leq b(B^{\top}x)$ almost surely, where $b(B^{\top}x)$ is a measurable function satisfied $E(b^2(B^{\top}X))<\infty$.
\item[A4] The density function $p(\cdot)$ of $B^{\top}X$  exists with support $\mathbb{C}$ and has
a continuous and bounded second-order derivative on the support
$\mathbb{C}$. The density $p(\cdot)$ satisfies
    \begin{eqnarray*}
    0<\inf_{B^{\top}X \in \mathbb{C}} p(B^{\top}X) \leq \sup_{B^{\top}X \in \mathbb{C}}p(B^{\top}X) < \infty.
     \end{eqnarray*}
\item[A5] For some positive integer $r$, the $r$th derivative of $g(\cdot)$ is bounded.
\item[A6] $Q(\cdot)$ is a bounded, symmetric and twice continuously differentiable kernel function such that $\int Q(u)du = 1$, $\int{u^{i}Q(u)du}=0$ and $\int u^rQ(u)du \neq 0$ for  $0 < i < r$, where $i$ is a nonnegative integer and $r$ is given by Condition A5.
\item[A7] $K(\cdot)$ is a bounded, symmetric and twice continuously differentiable kernel function satisfying $\int K(u)du = 1$.

\item[A8] $n \rightarrow \infty$, $h_1 \rightarrow 0$, $h \rightarrow 0$,
\begin{itemize}
     \item[1)] under the null or local alternative hypotheses,
         $nh^{q_1}\rightarrow \infty$, $h_1= O(n^{-\frac{1}{4+{q_1}}})$, $nh_1\rightarrow \infty$ and $nh^{{q_1}/2}h^{4r}_1 \rightarrow 0$;
      \item[2)] under global alternative hypothesis $H_{1}$, $nh^{q}\rightarrow \infty$, $h_1= O(n^{-1/(4+q)})$ and $nh^{q/2}h^{4r}_1 \rightarrow 0$,
\end{itemize}
where $\eta$ is given by Condition A6.
\end{itemize}

\begin{remark}
It is needed for DEE to assume Conditions A1 and A2. Under the linearity
condition and constant conditional variance condition, $DEE_{SIR}$ satisfies
Conditions A1 and A2. See Zhu et al (2010).
Conditions A3, A4, A5 and A6 are widely used for nonparametric estimation in the literature and are also needed for obtaining uniform convergence of $\hat{p}(\cdot)$ and $\hat{g}(\cdot)$.
Conditions A4 and A7 guarantee the asymptotic normality of our test statistic.
Applying a higher order kernel in A6 guarantees that the estimator $\hat{g}$ and $\hat{\sigma}^2 $ have sufficiently small biases, respectively,  see Powell et. al. (1989) and Hall and Marron (1990). To be specific, $\hat{\sigma}^2 $ has a convergence rate as $O_p(h^r_1)$.
Note that the density estimator $\hat{p}(\cdot)$ appears in the denominator of $\hat{g}(\cdot)$ and  small values of $\hat{p}(\cdot)$ may cause the estimator $\hat{g}(\cdot)$ and then the test statistic to be ill-behaved. Thus, Condition A4 can evade this problem. Thus, Conditions A4, A5 and A6 are needed for the test to be well-behaved.  Condition A8 is  similar to that in Fan and Li (1996), which was originally for model checking about the mean regression. We note that Zheng (2009) used a single bandwidth.  Actually, in our case, we could also use a single bandwidth. However, we found that  when we  respectively use different bandwidths $h$ and $h_1$ for  estimating the mean function $\hat{g}$ and constructing  the test statistic $T_n$,  the final test statistic is less sensitive to the bandwidth selection.  This phenomenon has been discussed in the literature such as Stute and Zhu (2005) pointing out that the optimal bandwidth for estimation is different from that for test statistic construction.
\end{remark}

\subsection{Proofs of the theorems}

\textbf{Proof of Theorem~\ref{theorem0}.} Under the assumptions designed in Zhu et. al. (2010),  their Theorem 2 shows that $M_{n}-M=O_p(n^{-1/2})$. Following the similar arguments used in Zhu and Ng (1995) or Zhu and Fang (1996), it is proved that  the root-$n$ consistency of the eigenvalues of $M_{n}$, namely, $\hat{\lambda}_{i}- \lambda_{i} =O_p(n^{-1/2})$.

Prove (i). It is obvious that under $H_0$, for any $l$ with $ 1< l \leq q_1$, $\lambda_{l}>0$. Therefore, we have  $\hat{\lambda}^2_{l}=\lambda^2_{l}+O_p(1/\sqrt{n})$. Since for any  $l$ with $q_1 < l \leq p$, $\lambda_{l}=0$, we have $\hat{\lambda}^2_{l}=O_p(1/n) = O_p(1/n)$.
For any $l < q_1$, we have $\lambda_{l}>0$, $\lambda_{l+1}>0$ and
\begin{eqnarray*}
\frac{\hat{\lambda}^2_{(q_1+1)}+c_n}{\hat{\lambda}^2_{q_1}+c_n}-\frac{\hat{\lambda}^2_{(l+1)}+c_n}{\hat{\lambda}^2_{l}+c_n}
&=&\frac{\lambda^2_{(q_1+1)}+c_n+O_p(1/n)}{\lambda^2_{q_1}+c_n+O_p(1/\sqrt{n})}
-\frac{\lambda^2_{(l+1)}+c_n+O_p(1/\sqrt{n})}{\lambda^2_{l}+c_n+O_p(1/\sqrt{n})}\\
&=&\frac{c_n+O_p(1/n)}{\lambda^2_{q_1}+c_n+O_p(1/\sqrt{n})}
-\frac{\lambda^2_{(l+1)}+c_n+O_p(1/\sqrt{n})}{\lambda^2_{l}+c_n+O_p(1/\sqrt{n})}.
\end{eqnarray*}
Taking  $\frac{\log{n}}{n} \leq c_n \rightarrow 0$, we can obtain
\begin{eqnarray*}
\frac{\hat{\lambda}^2_{(q_1+1)}+c_n}{\hat{\lambda}^2_{q_1}+c_n}
-\frac{\hat{\lambda}^2_{(l+1)}+c_n}{\hat{\lambda}^2_{l}+c_n}
\rightarrow  \frac{0}{\lambda^2_{q_1}}-\frac{\lambda^2_{(l+1)}}{\lambda^2_{l}}
=  -\frac{\lambda^2_{(l+1)}}{\lambda^2_{l}} < 0.
\end{eqnarray*}
Further, since for any $l>q_1$, we have $\lambda_{l}=0$ and $\lambda^2_{q_1} >0$ , we have
\begin{eqnarray*}
\frac{\hat{\lambda}^2_{(q_1+1)}+c_n}{\hat{\lambda}^2_{q_1}+c_n}-\frac{\hat{\lambda}^2_{(l+1)}+c_n}{\hat{\lambda}^2_{l}+c_n}
&=&\frac{\lambda^2_{q_1+1}+c_n+O_p(1/n)}{\lambda^2_{q_1}+c_n + O_p(1/\sqrt{n})}-\frac{\lambda^2_{l+1}+c_n+O_p(1/n)}{\lambda^2_{l}+c_n+O_p(1/n)}\\
&= & \frac{c_n+o_p(c_n)}{\lambda^2_{q_1}+c_n+O_p(1/\sqrt{n})}-\frac{c_n+o_p(c_n)}{c_n+o_p(c_n)}\\
&\rightarrow & -1 < 0.
\end{eqnarray*}
Therefore, altogether, we can conclude that $\hat{q} \rightarrow q_1$.

For part (ii), by replacing $q_1$ by $q$, and  using the same arguments as the above  we can obtain $\hat{q} \rightarrow q $ in probability. \hfill$\Box$

\textbf{Proof of Theorem~\ref{theo3.1}. } For notational convenience, denote $z_i=B^{\top}x_i$
 $g_i = g(B^{\top}x_i)$, $\hat{g}_i =\hat{g}(\hat{B}^{\top}_{\hat{q}}x_i)$,  $\mu_i=(y_i-g_i)^2 - \sigma^2$,
 $\epsilon_i= y_i-g_i$, $K_{Bij} =  K(B^{\top}(x_i -x_j)/h)$. Under the null hypothesis, without loss of generality, write $B_1=B$.

Since $\hat{\mu}_i \equiv : (y_i-\hat{g}_i)^2 - \hat{\sigma}^2=\mu_i -2\epsilon_i(\hat{g}_i-g_i) + (\hat{g}_i-g_i)^2 -(\hat{\sigma}^2-\sigma^2)$, we decompose the term $S_n$ to be:
 \begin{eqnarray*}
S_n &=& \frac{1}{n(n-1)}\sum_{i=1}^n\sum_{i \neq j}\frac{1}{h^{q_1}}K_{\hat{B}ij} \mu_i\mu_j +4 \frac{1}{n(n-1)}\sum_{i=1}^n\sum_{i \neq j}\frac{1}{h^{q_1}}K_{\hat{B}ij} \epsilon_i\epsilon_j(\hat{g}_i-g_i)(\hat{g}_j-g_j)\\
&&+\frac{1}{n(n-1)}\sum_{i=1}^n\sum_{i \neq j}\frac{1}{h^{q_1}}K_{\hat{B}_{\hat{q}}ij} (\hat{g}_i-g_i)^2(\hat{g}_j-g_j)^2
+\frac{1}{n(n-1)}\sum_{i=1}^n\sum_{i \neq j}\frac{1}{h^{q_1}}K_{\hat{B}_{\hat{q}}ij} (\hat{\sigma}^2-\sigma^2)\\
&&-4 \frac{1}{n(n-1)}\sum_{i=1}^n\sum_{i \neq j}\frac{1}{h^{q_1}}K_{\hat{B}_{\hat{q}}ij} \mu_i\epsilon_j(\hat{g}_j-g_j)
+2 \frac{1}{n(n-1)}\sum_{i=1}^n\sum_{i \neq j}\frac{1}{h^{q_1}}K_{\hat{B}_{\hat{q}}ij} \mu_i(\hat{g}_j-g_j)^2\\
&&-2 \frac{1}{n(n-1)}\sum_{i=1}^n\sum_{i \neq j}\frac{1}{h^{q_1}}K_{\hat{B}_{\hat{q}}ij} \mu_i(\hat{\sigma}^2_j-\sigma^2_j)
-4 \frac{1}{n(n-1)}\sum_{i=1}^n\sum_{i \neq j}\frac{1}{h^{q_1}}K_{\hat{B}_{\hat{q}}ij} \epsilon_i(\hat{g}_i-g_i)(\hat{g}_j-g_j)^2\\
&&4 \frac{1}{n(n-1)}\sum_{i=1}^n\sum_{i \neq j}\frac{1}{h^{q_1}}K_{\hat{B}_{\hat{q}}ij} \epsilon_i(\hat{g}_i-g_i)(\hat{\sigma}^2_j-\sigma^2_j)\\
&&-2 \frac{1}{n(n-1)}\sum_{i=1}^n\sum_{i \neq j}\frac{1}{h^{q_1}}K_{\hat{B}_{\hat{q}}ij} \epsilon_i(\hat{g}_i-g_i)^2(\hat{\sigma}^2-\sigma^2)+ o_p(n^{-1}h^{-{q_1}/2})\\
&\equiv :& \sum_{i=1}^{10} Q_{in} + o_p(n^{-1}h^{-{q_1}/2}).
 \end{eqnarray*}
The final equation is derived by applying Lemma 2 of Guo et. al (2014), where $\hat{q}=q_1$. We now deal with the terms.
First, consider the term $Q_{1n}$. By  Taylor expansion for $Q_{1n}$ with respect to $B$, we have
\begin{eqnarray*}
Q_{1n} \equiv : Q_{11n} + Q_{12n}  + Q_{13n} ,
  \end{eqnarray*}
where $Q_{11n}$,  $Q_{12n}$   and  $Q_{13n}$
have following forms:
\begin{eqnarray*}
Q_{11n} &=&\frac{1}{n(n-1)}\sum_{i=1}^n\sum_{i \neq j}\frac{1}{h^{q_1}}K_{Bij} \mu_i\mu_j ,\\
Q_{12n} &=&\frac{1}{n(n-1)}\sum_{i=1}^n\sum_{i \neq j}\frac{1}{h^{2{q_1}}}K'_{Bij} \mu_i\mu_j(\hat{B}_{\hat{q}}-B)^{\top}(x_i-x_j)
  \end{eqnarray*}
and
\begin{eqnarray*}
Q_{13n} &=&\frac{1}{n(n-1)}\sum_{i=1}^n\sum_{i \neq j}\frac{1}{h^{3{q_1}}}K''_{\tilde{B}ij} \mu_i\mu_j(\hat{B}_{\hat{q}}-B)^{\top}(x_i-x_j)(x_i-x_j)^{\top}(\hat{B}_{\hat{q}}-B).
  \end{eqnarray*}
where $\tilde{B}=\{\tilde{B}_{ij}\}_{p\times q_1}$ with  $\tilde{B}_{ij} \in [\min\{\hat{B}_{ij}, B_{ij}\}, \max\{\hat{B}_{ij}, B_{ij}\}]$.
Due to the two facts that $||\hat{B}_{\hat{q}}-B||=O_p(1/\sqrt{n})$ and the second-order differential function of $K_{B}(\cdot)$ is a bounded continuous function of $B$, we assert that replacing $\tilde{B}$ by $\hat{B}_{\hat{q}}$ does not affect the convergence rate of $Q_{13n}$.

By Theorem 1 in  Zheng (2009),  we obtain that:
\begin{eqnarray*}
nh^{{q_1}/2}Q_{11n} \to N(0, s^2).
\end{eqnarray*}
Since  $E(\mu_i)=0$, we have $E(Q_{21n})=0$. Then we compute the second order moment of $Q_{12n}$ as follows:
\begin{eqnarray*}
E(Q^2_{12n})&=&E\big{[}\frac{1}{n(n-1)}\sum_{i=1}^n\sum_{i \neq j}\frac{1}{h^{2{q_1}}}K'_{Bij} \mu_i\mu_j(\hat{B}_{\hat{q}}-B)^{\top}(x_i-x_j)\big{]}^2\\
&=&E\big{[}\frac{1}{n^2(n-1)^2}\frac{1}{h^{4{q_1}}}\sum_{i=1}^n\sum_{i' \neq j'}\sum_{i=1}^n\sum_{i' \neq j'}K'_{Bij}K'_{Bi'j'} \\
 &&\mu_i\mu_j\mu_{i'}\mu_{j'}(\hat{B}_{\hat{q}}-B)^{\top}(x_i-x_j) (x_{i'}-x_{j'})^{\top}(\hat{B}_{\hat{q}}-B)\big{]}
\end{eqnarray*}
Noting that $E(\mu_i\mu_j\mu_{i'}\mu_{j'}) \neq 0$ only if $i = i'$, $j = j'$ or $i = j'$, $j = i'$, we have
\begin{eqnarray*}
E(Q^2_{12n})&=&\frac{n(n-1)}{n^2(n-1)^2}\frac{1}{h^{4{q_1}}}E\big{[}(K'_{Bij})^2 \mu^2_i\mu^2_j(\hat{B}_{\hat{q}}-B)^{\top}(X_i-X_j) (x_{i}-x_{j})^{\top}(\hat{B}_{\hat{q}}-B)\big{]}\\
&=&\frac{1}{n(n-1)}\frac{1}{h^{4{q_1}}}E\big{[}(K'_{Bij})^2 \mu^2_i\mu^2_j(\hat{B}_{\hat{q}}-B)^{\top}(x_i-x_j) (x_{i}-x_{j})^{\top}(\hat{B}_{\hat{q}}-B)\big{]}\\
&=&\frac{1}{n(n-1)}\frac{1}{h^{4{q_1}}}\{E(\mu^2_i)\}^2(\hat{B}_{\hat{q}}-B)^{\top}E\big{\{}(K'_{Bij})^2 (x_i-x_j) (x_{i}-x_{j})^{\top}\big{\}}(\hat{B}_{\hat{q}}-B).
\end{eqnarray*}
By a variable transformation as
$u_1=(x_i- x_j)/h$, the above value is as
\begin{eqnarray*}
E(Q^2_{12n})&=&\frac{1}{n(n-1)}\frac{1}{h^{4{q_1}}}\{E(\mu^2_i)\}^2\int\int(K'_{Bij})^2(\hat{B}_{\hat{q}}-B)^{\top} (x_i-x_j)\\
&& (x_{i}-x_{j})^{\top}(\hat{B}_{\hat{q}}-B)p(B^{\top}x_i)p(B^{\top}x_j)dx_idx_j\\
&=&\frac{1}{n(n-1)}\frac{1}{h^{q_1}}\{E(\mu^2_i)\}^2\int\int(K'(u))^2
(\hat{B}_{\hat{q}}-B)^{\top}uu^{\top}(\hat{B}_{\hat{q}}-B)\\
&&p(B^{\top}x_i)p(B^{\top}(x_i-hu))dx_idu.
\end{eqnarray*}
By Taylor expansion of $p(B^{\top}(x_i-hu))$ about $x_i$ and Conditions A3-A7 in Appendix, we have
\begin{eqnarray*}
E(Q^2_{12n}) &=&\frac{1}{n(n-1)}\frac{1}{h^{q_1}}\{E(\mu^2_i)\}^2\int\int(K'(u))^2
(\hat{B}_{\hat{q}}-B)^{\top}uu^{\top}(\hat{B}_{\hat{q}}-B)\\
&&p(B^{\top}x_i)p(B^{\top}(x_i-hu))dx_idu\\
&=&\frac{1}{n(n-1)}\frac{1}{h^{q_1}}\{E(\mu^2_i)\}^2\int\int(K'(u))^2
(\hat{B}_{\hat{q}}-B)^{\top}uu^{\top}(\hat{B}_{\hat{q}}-B)\\
&&(p^2(B^{\top}x_i)+ p(B^{\top}x_i)p'(B^{\top}x_i)h^pu)dx_idu+o_p(\frac{1}{n(n-1)})\\
&=&\frac{1}{n(n-1)}\frac{1}{h^{q_1}}\{E(\mu^2_i)\}^2\int\int(K'(u))^2
(\hat{B}_{\hat{q}}-B)^{\top}uu^{\top}(\hat{B}_{\hat{q}}-B)p^2(B^{\top}x_i)dx_idu\\
&&+\frac{1}{n(n-1)}\frac{1}{h^{q_1}}\{E(\mu^2_i)\}^2\int\int(K'(u))^2
(\hat{B}_{\hat{q}}-B)^{\top}uu^{\top}(\hat{B}_{\hat{q}}-B)\\
&&p(B^{\top}x_i)p'(B^{\top}x_i)h^{q_1}u)dx_idu+o_p(\frac{1}{n(n-1)})O(\frac{1}{n})\\
&=&O_p(\frac{1}{n^2(n-1)h^{q_1}}).
\end{eqnarray*}
The application of  Chebyshiev's inequality yields that $|Q_{12n}|=o_p(n^{-1}h^{-{q_1}/2})$. Similarly, we can prove the term $Q_{13n}$ to have the rate: $Q_{13n}=o_p(n^{-1}h^{-{q_1}/2})$. Therefore, the above decomposition term $Q_{1n}$ convergences to a normal distribution:
\begin{eqnarray*}
nh^{{q_1}/2}Q_{1n}\stackrel{\mathrm{d}}{\rightarrow} N(0,s^2_1).
\end{eqnarray*}
To obtain the results of the theorem, it remains to prove that
$nh^{{q_1}/2}Q_{in}= o_p(1)$, $i=2, 3, \cdots, 10$.

Second, we consider the term $Q_{2n}$. Since
\begin{eqnarray*}
\frac{Q_{2n}}{4}&=&\frac{1}{n(n-1)}\sum_{i=1}^n\sum_{j \neq i}\frac{1}{h^{q_1}}K_{\hat{B}_{\hat{q}}ij}\varepsilon_i\varepsilon_j(\hat{g}_i-g_i)(\hat{g}_j-g_j)\frac{\hat{p}_{i}}{p_{i}}\frac{\hat{p}_{j}}{p_{j}}\\
&&+\frac{1}{n(n-1)}\sum_{i=1}^n\sum_{j \neq i}\frac{1}{h^{q_1}}K_{\hat{B}_{\hat{q}}ij}\varepsilon_i\varepsilon_j(\hat{g}_i-g_i)(\hat{g}_j-g_j)\left(\frac{\hat{p}_{i}-p_{i}}{p_{i}}\frac{\hat{p}_{j}-p_{j}}{p_{j}}-2\frac{(\hat{p}_{i}-p_{i})\hat{p}_{j}}{p_{i}p_{j}}\right)\\
&\equiv& \tilde{Q}_{2n} + o_p(\tilde{Q}_{2n}).
\end{eqnarray*}
Substituting the kernel estimates $\hat{g}$ and $\hat{p}$ into $\tilde{Q}_{2n}$, we have
\begin{eqnarray*}
\tilde{Q}_{2n}&=&\frac{1}{n^3(n-1)}\sum_{i=1}^n\sum_{j \neq i}\sum_{k=1}^n\sum_{l=1}^n\frac{1}{h^{q_1}h^{2{q_1}}_1}\frac{1}{p_{i}p_{j}}K_{\hat{B}_{\hat{q}}ij}Q_{\hat{B}_{\hat{q}}il}Q_{\hat{B}_{\hat{q}}jk}\epsilon_i\epsilon_j\\
&&\times(y_l-g(B^\top x_i))(y_k-g(B^\top x_j)).
\end{eqnarray*}
By the two order Taylor expansion for $\tilde{Q}_{2n}$ with respect to  $B$, we can have
\begin{eqnarray*}
\tilde{Q}_{2n}& \equiv &Q_{21n}+Q_{22n}+Q_{23n}
\end{eqnarray*}
where $Q_{21n}$, $Q_{22n}$ and $Q_{23n}$ have following forms:
\begin{eqnarray*}
Q_{21n}&=&\frac{1}{n^3(n-1)}\sum_{i=1}^n\sum_{j \neq i}\sum_{k=1}^n\sum_{l=1}^n\frac{1}{h^{q_1}h^{2{q_1}}_1}\frac{1}{p_{i}p_{j}}K_{Bij}Q_{Bil}Q_{Bjk}\varepsilon_i\varepsilon_j\\
&&(y_l-g(B^{\top}x_i))(y_k-g(B^{\top}x_j));
\end{eqnarray*}
\begin{eqnarray*}
Q_{22n}&=&\frac{1}{n^3(n-1)}\sum_{i=1}^n\sum_{j \neq i}\sum_{k=1}^n\sum_{l=1}^n\frac{1}{h^{q_1}h^{3{q_1}}_1}\frac{1}{p_{i}p_{j}}K_{Bij}Q'_{Bil}Q_{B jk}\varepsilon_i\varepsilon_j\\
&&(y_l-g(B^{\top}x_i))(y_k-g(B^{\top}x_j))(\hat{B}_{\hat{q}}-B)^{\top}(x_i-x_l)\\
&&+\frac{1}{n^3(n-1)}\sum_{i=1}^n\sum_{j \neq i}\sum_{k=1}^n\sum_{l=1}^n\frac{1}{h^{q_1}h^{3{q_1}}_1}\frac{1}{p_{i}p_{j}}K_{Bij}Q_{Bil}Q'_{B jk}\varepsilon_i\varepsilon_j\\
&&(y_l-g(B^{\top}x_i))(y_k-g(B^{\top}x_j))(\hat{B}_{\hat{q}}-B)^{\top}(x_j-x_k)\\
&&+\frac{1}{n^3(n-1)}\sum_{i=1}^n\sum_{j \neq i}\sum_{k=1}^n\sum_{l=1}^n\frac{1}{h^{q_1}h^{3{q_1}}_1}\frac{1}{p_{i}p_{j}}K'_{Bij}Q_{Bil}Q_{B jk}\varepsilon_i\varepsilon_j\\
&&(y_l-g(B^{\top}x_i))(y_k-g(B^{\top}x_j))(\hat{B}_{\hat{q}}-B)^{\top}(x_i-x_j)\\
&\equiv&(\hat{B}_{\hat{q}}-B)^{\top} Q_{221n}+(\hat{B}_{\hat{q}}-B)^{\top}Q_{222n}+(\hat{B}_{\hat{q}}-B)^{\top}Q_{223n};
\end{eqnarray*}
and
\begin{eqnarray*}
Q_{23n}&=&\frac{1}{n^3(n-1)}\frac{2}{h^{q_1}h^{4{q_1}}_1}\sum_{i=1}^n\sum_{j \neq i}\sum_{k=1}^n\sum_{l=1}^n\frac{1}{p_{1i}p_{1j}}K_{\tilde{B}ij}Q'_{\tilde{B} il}Q'_{\tilde{B} jk}\varepsilon_i\varepsilon_j\\	
&&(y_l-B^{\top}x_i))(y_k-g(B^{\top}x_j))(\hat{B}_{\hat{q}}-B)^\top(x_i-X_l)(x_j-x_k)^\top(\hat{B}_{\hat{q}}-B)\\
&&+\frac{1}{n^3(n-1)}\frac{2}{h^{q_1}h^{4{q_1}}_1}\sum_{i=1}^n\sum_{j \neq i}\sum_{k=1}^n\sum_{l=1}^n\frac{1}{p_{i}p_{j}}K'_{\tilde{B}ij}Q'_{\tilde{B} il}Q_{\tilde{B} jk}\varepsilon_i\varepsilon_j\\
&&(y_l-B^{\top}x_i))(y_k-g(B^{\top}x_j))(\hat{B}_{\hat{q}}-B)^\top(x_i-x_j)(x_i-x_l)^\top(\hat{B}_{\hat{q}}-B)\\
&&+\frac{1}{n^3(n-1)}\frac{2}{h^{q_1}h^{4{q_1}}_1}\sum_{i=1}^n\sum_{j \neq i}\sum_{k=1}^n\sum_{l=1}^n\frac{1}{p_{i}p_{j}}K'_{\tilde{B}ij}Q_{\tilde{B} il}Q'_{\tilde{B} jk}\varepsilon_i\varepsilon_j\\
&&(y_l-B^{\top}x_i))(y_k-g(B^{\top}x_j))(\hat{B}_{\hat{q}}-B)^\top(x_i-x_j)(x_j-x_k)^\top(\hat{B}_{\hat{q}}-B)\\
&&+\frac{1}{n^3(n-1)}\frac{1}{h^{q_1}h^{4{q_1}}_1}\sum_{i=1}^n\sum_{j \neq i}\sum_{k=1}^n\sum_{l=1}^n\frac{1}{p_{1i}p_{1j}}K''_{\tilde{B}ij}Q_{\tilde{B} il}Q_{\tilde{B} jk}\varepsilon_i\varepsilon_j\\	
&&(y_l-B^{\top}x_i))(y_k-g(B^{\top}x_j))(\hat{B}_{\hat{q}}-B)^\top(x_i-x_j)(x_i-x_j)^\top(\hat{B}_{\hat{q}}-B)\\
&&+\frac{1}{n^3(n-1)}\frac{1}{h^{q_1}h^{4{q_1}}_1}\sum_{i=1}^n\sum_{j \neq i}\sum_{k=1}^n\sum_{l=1}^n\frac{1}{p_{1i}p_{1j}}K_{\tilde{B}ij}Q''_{\tilde{B} il}Q_{\tilde{B} jk}\varepsilon_i\varepsilon_j\\
&&(y_l-B^{\top}x_i))(y_k-g(B^{\top}x_j))(\hat{B}_{\hat{q}}-B)^\top(x_i-x_l)(x_i-x_l)^\top(\hat{B}_{\hat{q}}-B)\\
&&+\frac{1}{n^3(n-1)}\frac{1}{h^{q_1}h^{4{q_1}}_1}\sum_{i=1}^n\sum_{j \neq i}\sum_{k=1}^n\sum_{l=1}^n\frac{1}{p_{1i}p_{1j}}K_{\tilde{B}ij}Q_{\tilde{B} il}Q''_{\tilde{B} jk}\varepsilon_i\varepsilon_j\\
&&(y_l-B^{\top}x_i))(y_k-g(B^{\top}x_j))(\hat{B}_{\hat{q}}-B)^\top(x_j-x_k)(x_j-x_k)^\top(\hat{B}_{\hat{q}}-B)\\
&\equiv& (\hat{B}_{\hat{q}}-B)^\top (Q_{231n}+ Q_{232n}+Q_{233n}+ Q_{234n}+Q_{235n}+ Q_{236n})(\hat{B}_{\hat{q}}-B);
\end{eqnarray*}
and where $\tilde{B}=\{\tilde{B}_{ij}\}_{p\times q_1}$ with  $\tilde{B}_{ij} \in [\min\{\hat{B}_{ij}, B_{ij}\}, \max\{\hat{B}_{ij}, B_{ij}\}]$. As described
 for the term $Q_{23n}$, we also assert that that replacing $\tilde{B}$ by $\hat{B}_{\hat{q}}$ does not affect the convergence rate.

For the term  $\tilde{Q}_{2n}$, we first consider  $Q_{21n}$. Since for any fixed $\hat{B}_{\hat{q}}$, $E(Q_{21n})=0$, we compute its second order moment as follows:

\begin{eqnarray*}
E(Q^2_{21n})&=&E\big{[}\frac{1}{n^3(n-1)}\sum_{i=1}^n\sum_{j \neq i}\sum_{k=1}^n\sum_{l=1}^n\frac{1}{h^{q_1}h^{2{q_1}}_1}\frac{1}{p_{i}p_{j}}K_{Bij}Q_{Bil}Q_{Bjk}\varepsilon_i\varepsilon_j\\
&&(y_l-g(B^{\top}x_i))(y_k-g(B^{\top}x_j))\big{]}^2\\
&=&E\big{[}\frac{1}{n^6(n-1)^2}\sum_{i=1}^n\sum_{j \neq i}\sum_{k=1}^n\sum_{l=1}^n\sum_{i'=1}^n\sum_{j' \neq i'}\sum_{k'=1}^n\sum_{l'=1}^n\frac{1}{h^{2{q_1}}h^{4{q_1}}_1}\frac{1}{p_{i}p_{i'}p_{j}p_{j'}}\\
&&K_{Bij}Q_{B il}Q_{B jk}K_{Bi'j'}Q_{B i'l'}Q_{B j'k'}
\varepsilon_i\varepsilon_j\varepsilon_i'\varepsilon_j'(Y_l-g(B^{\top}X_i))\\
&&(y_{l'}-g(B^{\top}x_{i'}))(y_k-g(B^{\top}x_j))(y_{k'}-g(B^{\top}x_{j'}))\big{]}.
\end{eqnarray*}
Noting that $E(\varepsilon_i\varepsilon_j\varepsilon_{i'}\varepsilon_{j'}) \neq 0$ only if $i = i'$, $j = j'$ or $i = j'$, $j = i'$, we have
\begin{eqnarray*}
&&E(Q^2_{21n})\\
&=&\frac{1}{n^6(n-1)^2h^{2{q_1}}h^{4{q_1}}_1}n(n - 1)(n - 2)^2(n - 3)^2E(\frac{1}{p^2_{i}}\frac{1}{p^2_{j}}K^2_{Bij}Q_{Bil}Q_{B jk}Q_{Bil'}Q_{B jk'}\\
&&(g_l - g_i)(g_k - g_j)(g_{l'} - g_i)(g_{k'} - g_j)\delta^4 +o((n^2h)^{-q}).
\end{eqnarray*}

By transforming variables as
$u_1=(z_i- z_j)/h$, $u_2=(z_i- z_l)/h_1$, $u_3=(z_j- z_k)/h_1$, $u_4=(z_i- z_{l'})/h_1$ and $u_5=(z_j- z_{k'})/h_1$, we can  have
\begin{eqnarray*}
E(Q^2_{21n})&=&\frac{h^{q_1}h^{4{q_1}}_1}{n^6(n-1)^2h^{2{q_1}}h^{4{q_1}}_1}n(n - 1)(n - 2)^2(n -3)^2\\
&&\int\int\int\int\int\int\frac{1}{p^2(Z_i)p^2(Z_i-hu_1)}K^2_{Bij}(u_1)Q_{B il}(u_2)Q_{B jk}(u_3)Q_{Bil'}(u_4)\\
&&Q_{Bjk'}(u_5)[g(z_i-h_1u_2)-g(z_i)][g(z_i-hu_1-h_1u_3)-g(z_i-hu_1]\\
&&[g(z_i-h_1u_4)-g(Z_i)][g(z_i-hu_1-h_1u_5)-g(z_i-hu_1]p(z_i)p(z_i-hu_1)\\
&&p_1(z_i-h_1u_2)p_1(z_i-hu_1-h_1u_3)p_1(z_i-h_1u_4)p_1(z_i-hu_1-h_1u_5)\\
&&dz_idu_1du_2du_3du_4du_5+o((n^2h^q)^{-1}).
\end{eqnarray*}

By taking Taylor expansions of $g(z_i - hu_1) - g(z_i)$ and similar terms at $z_i$
and using Conditions A4, A5 and A6, we have
$$E(Q^2_{21n})=O_p(\frac{h^{q_1}h^{4{q_1}}_1h^{4r}_1}{n^2h^{2{q_1}}h^{4{q_1}}_1})=O_p(\frac{h^{4r}_1}{n^2h^{q_1}}).$$
the application of  Chebyshiev's inequality leads to $|Q_{21n}|=o_p(1/(nh^{{q_1}/2}))$.

Similarly, the terms $Q_{22in}$ and $Q_{23jn}$ for $\{i=1,2,3, j=1,\cdots, 6\}$ can be proved to have the following rates:
$Q_{22in}=o_p(\frac{h^{r}_1}{nh^{{q_1}/2}}\cdot\frac{1}{h^{q_1/2}_1})=o_p(\frac{h^{r}_1}{nh^{{q_1/2}}h^{q_1/2}_1})$ and
$Q_{23jn}=o_p(\frac{h^{r/2}_1}{nh^{{q_1}/2}}\cdot\frac{1}{h^{q_1}_1})=o_p(\frac{h^{r/2}_1}{nh^{{q_1}/2}h^{q_1}_1})$.
As  $h_1= O(n^{-1/(4+q_1)})$, we can get
$|Q_{22n}|=o_p(\frac{h^{r}_1}{nh^{{q_1/2}}h^{q_1/2}_1}\cdot \frac{1}{\sqrt{n}}) = o_p(1/(nh^{{q_1}/2}))$ and $|Q_{23n}|=o_p(\frac{h^{r/2}_1}{nh^{{q_1}/2}h^{q_1}_1}\cdot \frac{1}{n}) = o_p(1/(nh^{{q_1}/2}))$.
Thus we arrive at the result that $nh^{{q_1}/2}Q_{2n} = o_p(1)$.

Now we consider $Q_{3n}$. Following the similar argument for proving Theorem 3 of Collomb and H\"{a}rdle (1986), we have
\begin{eqnarray*}
\sup_{t\in C^q}|\hat{g}(t)-g(t)|=O_p\Big (\sqrt{\frac{\ln{n}}{nh^{q_1}_1}}\Big ).
\end{eqnarray*}
Further,
\begin{eqnarray*}
Q_{3n}&\leq&\frac{1}{n(n-1)}\sum_{i=1}^n\sum_{j \neq i}\frac{1}{h^{q_1}}K_{Bij}(\sup_{t \in C^q}|\hat{g}_i(t)-g_i(t)|^4)\\
&=&O_p\Big (\frac{\ln^2{n}}{n^2h^{2q_1}_1}\Big )=o_p(\frac{1}{nh^{{q_1}/2}}),
\end{eqnarray*}
we can obtain $nh^{{q_1}/2}Q_{3n} = o_p(1)$.

Similarly as  Hall and Marron (1990), we can easily obtain that under the null hypothesis,  $\hat{\sigma}^2 =\sigma^2 +O_p(h^{2r}_1)$.
Since $nh^{{q_1}/2}h^{4r}_1 \rightarrow 0$, we have $Q_{4n} = O_p(h^{4r}_1) = o_p((nh^{{q_1}/2})^{-1})$.
Using the same argument as the above, we can prove
$nh^{{q_1}/2}Q_{5n}=o_p(1), \cdots ,nh^{{q_1}/2}Q_{10n}=o_p(1)$.
Hence, we can conclude that
 $$nh^{{q_1}/2}S_{1n}\stackrel{\mathrm{d}}{\rightarrow} N(0,s^2).$$
Second, we also need to prove $\hat{s}^2 \stackrel{\mathrm{p}}{\rightarrow} s^2$. Note that $\hat{\beta}$, $\hat{B}_{\hat{q}}$ and $\hat{g}$ are respectively the uniform consistency estimators of $\beta$, $B$ and $g$. Thus,
\begin{eqnarray*}
\hat{s}^2= \frac{2}{n(n-1)}\sum_{i=1}^n\sum_{j \neq
i}\frac{1}{h^{q_1}}K^2_{Bij}(\varepsilon^2_i-\sigma^2)^2(\varepsilon^2_i-\sigma^2)^2+o_p(1)\equiv:
s_n+o_p(1),
\end{eqnarray*}
where $s_n$ is an $U$-statistic with the kernel as:
\begin{eqnarray*}
H_{n}(w_i,w_j)=\frac{1}{h^{q_1}}K^2_{Bij}(\varepsilon^2_i-\sigma^2)^2(\varepsilon^2_j-\sigma^2)^2,
\end{eqnarray*}
with $w_i=(x_i, \varepsilon^2_i)$ for $i=1,\cdots, n.$
It can be computed to obtain that
\begin{eqnarray*}
E(H_{n}(w_i,w_j))&=&\int\int{}\frac{1}{h^{q_1}}K^2(\frac{z_i-z_j}{h})Var(\varepsilon^2_i|z_i)Var(\varepsilon^2_j|z_j)p(z_i)p(z_j)dz_idz_j\\
&=&\frac{1}{h^{q_1}}\int\int{}K^2(u)Var(\varepsilon^2_i|z_i)Var(\varepsilon^2_j|z_i-hu)p(z_i)p(z_i-hu)h^{q_1}dz_idu\\
&=&\int{}K^2(u)du\int{}[Var(\varepsilon^2_i|z_i)]^2p^2(z_i)dz_i+o_p(1)=s^2_1+o_p(1).
\end{eqnarray*}
Here  the  variable transformation  $u=(z_i- z_j)/h$ is used.
Using the similar argument used to prove Lemma 3.1 of  Zheng (1996), we have
 $s_{n}=
E(H_{1n}(w_i,w_j))+o_p(1) = s^2 +o_p(1)$. Thus,
$$\hat{s}^2_1 \stackrel{\mathrm{p}}{\rightarrow} s^2.$$
Finally, Slutsky lemma is applied to detain
\begin{eqnarray*}
T_{n} \stackrel{\mathrm{d}}{\rightarrow} N(0,1).
\end{eqnarray*}
The proof of Theorem~\ref{theo3.1} is concluded.  \hfill$\Box$

\textbf{Proof of Lemma~\ref{lemma1}}.
  Consider MRRE when SIR-based DEE is used.  To derive  $M_{n}-M = O_p(C_n)$, we only need to prove that $M_n(t)- M(t) = O_p(C_n)$ uniformly,  where
$M(t) = \Sigma^{-1}Var(E(X|I(Y\leq t))=\Sigma^{-1}(\nu_1-\nu_0)(\nu_1-\nu_0)^{\top}p_t(1-p_t)$, $\Sigma$ is the covariance matrix of $X$, $\nu_0 = E(X|I(Y\leq t)=0)$, $\nu_0 = E(X|I(Y\leq t)=1)$ and $p_t=E(I(Y\leq t))$.
It is easy to see that
\begin{eqnarray*}
\nu_{1}-\nu_{0} &=&\frac{E(XI(Y \leq t))}{p_t}-\frac{E(XI(Y > t))}{1-p_t}\\
&=& \frac{E(XI(Y \leq t)-E(X))E(I(Y \leq t))}{p_t(1-p_t)}.
\end{eqnarray*}
Thus, $M(t)$ can also be rewritten as
\begin{eqnarray*}
M(t)  &=& \Sigma^{-1}[E\{(X-E(X))I(Y \leq t)\}][E\{(X-E(X))I(Y \leq t)\}]^\top\\
&=:&\Sigma^{-1} m(t)m(t)^{\top}
\end{eqnarray*}
where $ m(t)= E\{(X-E(X))I(Y \leq t)\}$. Therefore,  $m(t)$ can be estimated by:
\begin{eqnarray*}
m_{n}(t) = n^{-1}\sum_{i=1}^n (x_{i}- \bar{x})I(y_{i} \leq t),
\end{eqnarray*}
and then $M(t)$ can be estimated by
$$M_{n}(t) =\hat{\Sigma}^{-1}L_{n}(t), $$
where $ \bar{x}=\frac{1}{n}\sum_{i=1}^nx_i$, $L_{n}(t)=m_{n}(t)m_{n}(t)^{\top}$ and $\hat{\Sigma}$ is the sample version of $\Sigma$.\\
Since the response under the local alternative is related to $n$, we write the response under the null and local alternative hypotheses as $Y$ and $Y_n$ respectively.
Further, it is  noted that:
\begin{eqnarray*}
E\{XI(Y_{n} \leq t)\}-E\{XI(Y \leq t)\} =E[X\{P(Y_{n} \leq t|X)\}]-E[X\{P(Y \leq t|X)\}].
\end{eqnarray*}
Under $H_{1n}$, because $Var(\varepsilon|X)= \sigma^2 +C_nf(B^{\top}X)$, we rewrite the local alternative model as  $Y=g(B^{\top}_1X)+\varepsilon(1+C_nf(B^{\top}X)/2).$ Thus, we have for all $t$,
\begin{eqnarray*}
&&P(Y_{n} \leq t|X)-P(Y \leq t|X)\\
&=&P(\varepsilon \leq \frac{t-g(B^{\top}_1X)}{1+C_nf(B^{\top}X)/2}|X)-P(\varepsilon \leq t-g(B^{\top}_1X)|X)\\
&=&P(\varepsilon \leq t-g(B^{\top}_1X)+C_n(t-g(B^{\top}_1X))f(B^{\top}X)/2|X)\\
&&-P(\varepsilon \leq t-g(B^{\top}_1X)|X)+o_p(C_n)\\
&=&F_{Y|X}(t-C_n(t-g(B^{\top}_1X))f(B^{\top}X)/2)-F_{Y|X}(t)\\
&=&-C_n(t-g(B^{\top}_1X))f(B^{\top}X)/2f_{Y|X}(t)+o_p(C_n).
\end{eqnarray*}
Therefore, under  Condition A2,  we can conclude that
\begin{eqnarray*}
&&n^{-1}\sum_{i=1}^n x_{i}I(y_{ni} \leq t)-E\{XI(Y \leq t)\}\\
&=&n^{-1}\sum_{i=1}^n x_{i}I(y_{ni} \leq t)-E\{XI(Y_n \leq t)\}+\{E\{XI(Y_n \leq t)-E\{XI(Y \leq t)\}\\
&=&O_p(\max{(C_n, n^{-1/2})}).
\end{eqnarray*}
Using the similar arguments used for proving Theorem 3.2 of Li et al. (2008), we can derive that  $M_n(t)-M(t) = O_p(\max{(C_n, n^{-1/2})})$ uniformly. Thus,   $M_{n}-M = O_p(\max{(C_n, n^{-1/2})})$.

As Zhu and Fang (1996) and Zhu and Ng (1995) demonstrated, since $M_{n}-M = O_p(\max{(C_n, n^{-1/2})})$, we conclude $\hat{\lambda}_{i}- \lambda_{i} = O_p(\max{(C_n, n^{-1/2})})$ where $\hat{\lambda}_{p} \leq \hat{\lambda}_{(p-1)}\leq \cdots \leq \hat{\lambda}_{1}$ are the eigenvalues of the matrix $M_n$.

Note that under the null hypothesis, we have $\lambda_{p}= \cdots = \lambda_{p-q_1}=0$ and $0< \lambda_{q_1}\leq \cdots \leq \lambda_{1}$ to be the eigenvalues of the matrix $M$. Since $\frac{\log{n}}{nh^{q_1/2}} = c_n \rightarrow 0$ and under the local alternative hypotheses $H_{1n}$ with $C_n = 1/(n^{1/2}h^{q_1/4})$, we have $C^2_n=o_p(c_n)$.
 Similarly as the proof for Theorem ~\ref{theorem0}.
 It is clear that for any $l\leq q_1$, we have $\lambda_{1}>0$. Then we have $\hat{\lambda}^2_{1}=\lambda^2_1+O_p(C_n)$. On the other hand, for any $q_1 < l \leq p$, as  we have $\lambda_{l}=0$, $\hat{\lambda}^2_{l}= \lambda^2_l+O_p(C^2_n)=O_p(C^2_n)$.
When $l>q_1$,  MRRE  is computed to be
\begin{eqnarray*}
\frac{\hat{\lambda}^2_{q_1+1}+c_n}{\hat{\lambda}^2_{q_1}+c_n}-\frac{\hat{\lambda}^2_{(l+1)}+c_n}{\hat{\lambda}^2_{l}+c_n}
&=&\frac{\lambda^2_{q_1+1}+c_n+O_p(C^2_n)}{\lambda^2_{q_1}+c_n + O_p(C_n)}-\frac{\lambda^2_{l+1}+c_n+O_p(C^2_n)}{\lambda^2_{l}+c_n+O_p(C^2_n)}\\
&=&\frac{\lambda^2_{q_1+1}+c_n+o_p(c_n)}{\lambda^2_{q_1}+c_n + O_p(C^2_n)}-\frac{\lambda^2_{l+1}+c_n+o_p(c_n)}{\lambda^2_{l}+c_n+o_p(c_n)}\\
&=&\frac{c_n+o_p(c_n)}{\lambda^2_{q_1}+c_n + O_p(C^2_n)}-\frac{c_n+o_p(c_n)}{c_n+o_p(c_n)}.
\end{eqnarray*}
Thus, we have
\begin{eqnarray*}
\frac{\hat{\lambda}^2_{q_1+1}+c_n}{\hat{\lambda}^2_{q_1}+c_n}-\frac{\hat{\lambda}^2_{(l+1)}+c_n}{\hat{\lambda}^2_{l}+c_n}
\rightarrow  -1<0.
\end{eqnarray*}
When $1 \leq l < q_1$,  MRRE  is computed to be:
\begin{eqnarray*}
\frac{\hat{\lambda}^2_{q_1+1}+c_n}{\hat{\lambda}^2_{q_1}+c_n}-\frac{\hat{\lambda}^2_{(l+1)}+c_n}{\hat{\lambda}^2_{l}+c_n}
&=&\frac{\lambda^2_{q_1+1}+c_n+O_p(C^2_n)}{\lambda^2_{q_1}+c_n + O_p(C_n)}-\frac{\lambda^2_{l+1}+c_n+O_p(C_n)}{\lambda^2_{l}+c_n+O_p(C_n)}\\
&=&\frac{c_n+o_p(c_n)}{\lambda^2_{q_1}+c_n + O_p(C^2_n)}-\frac{\lambda^2_{l+1}+c_n+o_p(c_n)}{\lambda^2_{l}+c_n+o_p(c_n)}.
\end{eqnarray*}
Then
\begin{eqnarray*}
\frac{\hat{\lambda}^2_{q_1+1}+c_n}{\hat{\lambda}^2_{q_1}+c_n}-\frac{\hat{\lambda}^2_{(l+1)}+c_n}{\hat{\lambda}^2_{l}+c_n}
\rightarrow -\frac{\lambda^2_{l+1}}{\lambda^2_{l}} <0.
\end{eqnarray*}
Therefor, we can conclude that $\hat{q} \rightarrow q_1$. \hfill$\Box$

\textbf{Proof of Theorem~ \ref{theo2.2}.} First, we prove Part (I). Applying the same decomposition technique as that in Theorem \ref{theo3.1}, $S_{n}$ can be decomposed in the following form:
\begin{eqnarray*}
S_{n}&\equiv:& \sum_{i=1}^{10} Q_{in},
\end{eqnarray*}
where $\{Q_{in}\}_{i=1}^{10}$ is defined in Theorem \ref{theo3.1}.
Note that $\hat{B}_{\hat{q}}$, $\hat{g}$ and $\hat{\sigma}^2$ are  respectively uniform consistent estimators of $B$, $g$ and $\sigma^2$. Then we have $S_{n} = Q_{1n} + o_p(1)$. It is clear that  $Q_{1n}$ is a $U-$statistic with kernel
$H_n= \frac{1}{h^q}K_{Bij}u_iu_j$.
Denote $w_i=(x_i, \varepsilon^2_i)$ for $i=1,\cdots, n.$
Under the alternative hypothesis, due to the fact $E(u_i|B^{\top}x_i)=Var(\varepsilon^2_i|B^{\top}x_i)-Var(\varepsilon^2_i)$, we have
\begin{eqnarray*}
E[H_n(w_i,w_i)]&=&\int\int K_{h}(B^{\top}x_i-B^{\top}x_j)[Var(\varepsilon_i|B^{\top}x_i)-Var(\varepsilon_i)]\\
&&[Var(\varepsilon_j|B^{\top}x_j)-Var(\varepsilon_j)]p(B^{\top}x_i)p(B^{\top}x_j)dB^{\top}x_idB^{\top}x_j.
\end{eqnarray*}
using the transformed variable $u=(z_i- z_j)/h$, we have
\begin{eqnarray*}
E[H_n(Z_i,Z_j)]&=&\frac{1}{h^p}\int\int K(u)[Var(\varepsilon_i|B^{\top}x_i)-Var(\varepsilon_i)]\\
&&[Var(\varepsilon_j|B^{\top}x_i-hu)-Var(\varepsilon_j)]p(B^{\top}x_i)p(B^{\top}x_i-hu)dB^{\top}x_idu\\
&=&E\big{(}[Var(\varepsilon|B^{\top}x)-Var(\varepsilon)]^2p(B^{\top}x)\big{)}.
\end{eqnarray*}
Lemma 3.1 of Zheng (1996) yields that
$$S_n = Q_{1n} + o_p(1) = E[H_n(w_i,w_j)] +o_p(1)=
E\big{\{}[Var(\varepsilon|B^{\top}X)-Var(\varepsilon)]^2p(B^{\top}X)\big{\}}+o_p(1).$$
Similarly, we can prove that $\hat{s}^2 \stackrel{\mathrm{p}}{\rightarrow} s^2,$ and then
$$T_n/(nh^{\frac{q}{2}})\stackrel{\mathrm{d}}{\rightarrow}
E\big{\{}[Var(\varepsilon|B^{\top}X)-Var(\varepsilon)]^2p(B^{\top}X)\big{\}}/s.$$
Prove Part (II). Under the  local alternative hypotheses $H_{1n}$, similar arguments used for proving Theorem \ref{theo3.1}, we can show that $S_n = Q_{1n} + o_p((nh^{q_1})^{-1})$. Let $\varepsilon^2_{2i}=\varepsilon^2_{i}-C_nf(B^{\top}x_i)$. Under the local alternative, $E(\varepsilon^2_{2i}|x_i)=\sigma^2$. $Q_{1n}$ is then  decomposed as:
\begin{eqnarray*}
Q_{1n}&=&\frac{1}{n(n-1)}\sum_{i=1}^n\sum_{j \neq i}\frac{1}{h^{q_1}}K_{\hat{B}_{\hat{q}}ij}u_iu_j\\
&=& \frac{1}{n(n-1)}\sum_{i=1}^n\sum_{j \neq i}\frac{1}{h^{q_1}}K_{\hat{B}_{\hat{q}}ij}(\varepsilon^2_{2i}-\sigma^2)(\varepsilon^2_{2j}-\sigma^2)\}\\
&&+C_n\{\frac{1}{ n(n-1)}\sum_{i=1}^n\sum_{j \neq i}\frac{1}{h^{q_1}}K_{\hat{B}_{\hat{q}}ij}f(B^{\top}x_i)(\varepsilon^2_{2j}-\sigma^2)\}\\
&&+C^2_n\{\frac{1}{n(n-1)}\sum_{i=1}^n\sum_{j \neq i}\frac{1}{h^{q_1}}K_{\hat{B}_{\hat{q}}ij}f(B^{\top}x_i)f(B^{\top}x_j)\\
&\equiv& W_{1n}+C_nW_{2n}+C^2_nW_{3n}.
\end{eqnarray*}
$W_{1n}$ has the following decomposition by Taylor expansion:
\begin{eqnarray*}
W_{1n}&=&\frac{1}{n(n-1)}\sum_{i=1}^n\sum_{j \neq i}\frac{1}{h^{q_1}}K_{B_1ij}(\varepsilon^2_{2i}-\sigma^2)(\varepsilon^2_{2j}-\sigma^2)\\
&&+(\hat{B}_{\hat{q}}-B_1)^{\top}\big{\{}\frac{1}{n(n-1)}\sum_{i=1}^n\sum_{j \neq i}\frac{1}{h^{2{q_1}}}K'_{B_1ij}(\varepsilon^2_{2i}-\sigma^2)(\varepsilon^2_{2j}-\sigma^2)
(x_i-x_j)\}\big{\}}\\
&&+(\hat{B}_{\hat{q}}-B_1)^{\top}\big{\{}\frac{1}{n(n-1)}\sum_{i=1}^n\sum_{j \neq i}\frac{1}{h^{3{q_1}}}K''_{\tilde{B}ij}(\varepsilon^2_{2i}-\sigma^2)\\
&&(\varepsilon^2_{2j}-\sigma^2)(x_i-x_j) (x_{i'}-x_{j'})^{\top}\big{\}}(\hat{B}_{\hat{q}}-B_1)\\
&\equiv:& W_{11n}+ (\hat{B}_{\hat{q}}-B_1)^{\top}W_{12n} + (\hat{B}_{\hat{q}}-B_1)^{\top}W_{13n}(\hat{B}_{\hat{q}}-B_1),
\end{eqnarray*}
where $\tilde{B}=\{\tilde{B}_{ij}\}_{p\times q_1}$ with  $\tilde{B}_{ij} \in [\min\{\hat{B}_{1ij}, B_{ij}\}, \max\{\hat{B}_{ij}, B_{1ij}\}]$. Here for the terms $W_{12n}$ and $W_{12n}$, using the similar argument of terms $Q_{21n}$ and $Q_{23n}$ in Theorem~\ref{theo3.1}, respectively, we prove to have the following rates: $W_{22n}=o_p(\frac{1}{\sqrt{n}h^{{q_1}/2}})$, $W_{23n}=(\frac{1}{\sqrt{n}h^{{q_1}/2}})$.
On the other hand, in the same way as that for proving  Theorem 1 in Zheng (2009), we can easily derive that
$nh^{\frac{q_1}{2}}W_{11n}\stackrel{\mathrm{d}}{\rightarrow} N(0,s^2)$. Thus, we have
$$nh^{\frac{q_1}{2}}W_{1n}\stackrel{\mathrm{d}}{\rightarrow} N(0,s^2).$$
According to Lemma 3.1 of Zheng (1996), it is easy to prove that $\sqrt{n}W_{2n}=O_p(1)$. Thus, when $C_n = n^{-\frac{1}{2}}h^{-\frac{q_1}{4}}$, $nh^{\frac{q_1}{2}}W_{2n} = o_p(1)$.

Finally, consider the term $W_{3n}$. Also by Taylor expansion, we have
\begin{eqnarray*}
W_{3n}&=&\{\frac{1}{n(n-1)}\sum_{i=1}^n\sum_{j \neq i}\frac{1}{h^p}K_{B_1ij}f(B^{\top}x_i)f(B^{\top}x_j)\\
&&+(\hat{B}_{\hat{q}}-B_1)^{\top}\big{\{}\frac{1}{n(n-1)}\sum_{i=1}^n\sum_{j \neq i}\frac{1}{h^{2q}}K'_{\tilde{B}ij}f(B^{\top}x_i)f(B^{\top}x_j)
(x_i-x_j)\}\big{\}}\\
&\equiv:& W_{31n}+ (\hat{B}_{\hat{q}}-B_1)^{\top}W_{32n},
\end{eqnarray*}
where $\tilde{B}=\{\tilde{B}_{ij}\}_{p\times q_1}$ with  $\tilde{B}_{ij} \in [\min\{\hat{B}_{ij}, B_{1ij}\}, \max\{\hat{B}_{ij}, B_{1ij}\}]$. We can also asset that replacing $\tilde{B}$ by $B_1$ can not impact the converging rate of the term $W_{32n}$. Note that $W_{32n} $ can be written as an $U-$Statistic with the kernel:
\begin{eqnarray*}
H_{n}(x_i,x_j)=\frac{1}{h^{2{q_1}}}K'_{B_1ij}f(B^{\top}x_i)f(B^{\top}x_j)
(x_i-x_j)+ \frac{1}{h^{2{q_1}}}K'_{B_1ji}f(B^{\top}x_i)f(B^{\top}x_j)
(x_j-x_i)
\end{eqnarray*}
Using   $U-$Statistics theory (e. g.  Serfling 1980), we have $W_{32n} =O_p(1)$. Additionally, $W_{31n}$ is also an $U-$Statistic with the kernel:
\begin{eqnarray*}
H_{n}(w_i,w_j)=\frac{1}{h^{q_1}}K_{B_1ij}f(B^{\top}x_i)f(B^{\top}x_j),
\end{eqnarray*}
where  $w_i=(x_i, \varepsilon_i)$ for $i=1,\cdots, n.$
First, we compute the first moment of $H_{n}(w_i,w_j)$ as
\begin{eqnarray*}
E(H_{n}(w_i,w_j))=E\{\frac{1}{h^{q_1}}K_{B_1ij}E[f(B^{\top}x_i)|B^{\top}_1x_i]E[f(B^{\top}x_j)|B^{\top}_1x_j]\}.
\end{eqnarray*}
For notational convenience, we assume $M(B^{\top}_1x_i)=E[f(B^{\top}x_i)|B^{\top}_1x_i]$ and $z_{i}=B^{\top}_1x_i$. Further, $E[H_{n}(w_i,w_j)]$ can be computed as
\begin{eqnarray*}
E(H_{n}(w_i,w_j))&=&E\{\frac{1}{h^{q_1}}K_{B_1ij}M(B^{\top}_1x_i)M(B^{\top}_1x_j)\}\\
&=&\int\int\frac{1}{h^{q_1}}K(\frac{z_{i}-z_{j}}{h})
M(z_i)M(z_j)p(z_{i})p(z_{j})dz_{i}dz_{j}\\
&=&\int\int{}K(u)M(z_{i})M(z_{i}-hu)
p(z_{i})p(z_{i}-hu)dz_idu\\
&=&\int{}K(u)du\int{} M(z_{i})M(z_{i})p(z_{i})p(z_{i})dz_i +o_p(h)\\
&=& E\{[E\{f(B^{\top}X)|B^{\top}_1X\}]^2p(B^{\top}_1X)\},
\end{eqnarray*}
where $p(\cdot)$ denotes the density function of $B^{\top}_1X$.  Similarly, we have the consistency of $W_{3n}$ as it goes to $E\{[E\{f(B^{\top}X)|B^{\top}_1X\}]^2p(B^{\top}_1X)\}$ in probability.  Additionally, similarly as the above proof for Part (I) of Theorem \ref{theo3.1}, it is easy to prove $\hat{s}^2 \stackrel{\mathrm{p}}{\rightarrow} s^2$.

Thus, invoking Slutsky theorem, we can conclude that
$$T_{n}\stackrel{\mathrm{d}}{\rightarrow} N(E\{[E\{f(B^{\top}X)|B^{\top}_1X\}]^2p(B^{\top}_1X)\}/s, 1).$$ \hfill$\Box$

\newpage

\leftline{\large\bf References}

\begin{description}
\item Collomb. G and H\"{a}rdle, W.  (1986). Strong uniform convergence rates in robust nonparametric time series analysis and prediction:Kernel regression estimation from dependent observations. {\it Stochastic Processes and Applications}, {\bf 23}, 77-89.


\item Cook, R. D. (1998). {\it Regression Graphics: Ideas for Studying Regressions Through Graphics. } {New York: Wiley.}


\item Cook, R. D and Weisberg, S. (1983). Diagnostics for heteroscedasticity in regression,
{\it Biometrika}, {\bf70}, 1-10.

\item Cook, R. D. and Weisberg, S. (1991). Discussion of ¡°Sliced inverse regression for dimension reduction,¡± by K. C. Li. {\it Journal of the American Statistical Association.}, {\bf 86}, 316-342.

\item Dette, H. (2002). A consistent test for heteroscedasticity in nonparametric regression based on the kernel method, {\it Journal of Statistical Planning and Inference}, {\bf 103}, 311-329.


\item  Dette, H. and Munk, A. (1998). Testing heteroscedasticity in nonparametric regression,
{\it Journal of the Royal Statistical Society: Series B}, {\bf 60}, 693-708.

\item  Dette, H., Neumeyer, N. and Van Keilegom, I. (2007). A new test for the parametric form of the variance function in nonparametric regression, {\it Journal of the Royal Statistical Society: Series B}, {\bf 69(5)},  903-917.

\item Fan, Y. and  Li, Q., (1996). Consistent model specication tests: omitted variables and semiparametric functional forms. {\it Econometrica}, {\bf 64}, 865-890.

\item Guo, X. Wang, T. and Zhu, L. X. (2014). Model checking for generalized linear models: a
dimension-reduction model-adaptive approach. \url{http://arxiv.org/abs/1405.2134}

\item Hall, P. and  Marron, J. S. (1990). On Variance Estimation in Nonparametric
Regression, {\it Biometrika}, {\bf 77}, 415-419.



\item Li, B., Wen, S. Q. and Zhu, L. X. (2008). On a Projective Resampling method for dimension reduction with multivariate responses. {\it Journal of the American Statistical Association}. {\bf 103}, 1177-1186.

\item Liero, H. (2003). Testing homoscedasticity in nonparametric regression, {\it Journal of Nonparametric Statistics}, {\bf 15(1)}, 31-51.

\item Lin, J. G. and Qu, X. Y. (2012). A consistent test for heteroscedasticity in semi-parametric regression with nonparametric variance function based on the kernel method, {\it Statistics}, {\bf 46}, 565-576.


\item Naik, D. N. and Khattree, R. (1996). Revisiting Olympic track records: Some practical considerations in the principal component analysis. {\it The American Statistician} {\bf 50}, 140-144.

\item Powell, J. L., Stock, J. H. and Stoker, T. M. (1989) Semiparametric Estimation
of Index Coefficients, {\it Econometrica}, {\bf 57}, 1403-1430.

\item Serfling, R. J. (1980). {\it Approximation Theorems of Mathematical Statistics.}
John Wiley, New York.

\item Simonoff, J. S. and Tsai, C. L. (1994) Improved tests for nonconstant variance in regression based on the modified profile likelihood, {\it Journal of Applied Statistics}, {\bf 43}, 357-370.

\item Stute, W. and Zhu, L. X. (2005). Nonparametric checks for single-index models,
{ \it The Annals of Statistics}, {\bf 33}, 1048-1083.


\item Tsai, C. L. (1986). Score test for the first-order autoregressive model with heteroscedasticity,
{\it Biometrika}, {\bf 73}, 455-460.


\item Xia, Q., Xu, W. and Zhu. L. (2014). Consistently determining the number of factors in multivariate volatility modelling, {\it Statistica Sinica}, accepted.

\item Xia, Y. C., Tong, H., Li, W. K. and Zhu, L. X. (2002). An adaptive estimation of dimension reduction space. {\it Journal of the Royal Statistical Society: Series B}, {\bf 64}, 363-410.

\item  Zheng, J. X. (1996). A Consistent Test of Functional Form Via Nonparametric
Estimation Techniques, {\it Journal of Econometrics}, {\bf 75}, 263-289.

\item Zheng, J. X. (2009). Testing heteroscedasticity in nonlinear and nonparametric regressions, {\it Canadian Journal of Statistics}, {\bf 37}, 282-300.

\item Zhu, L. X. and Fang, K. T. (1996). Asymptotics for the kernel estimates of sliced inverse regression. {\it Annals of Statistics}, {\bf 24}, 1053-1067.

\item Zhu, L. X. and Ng, K. W. (1995). Asymptotics for sliced inverse regression. {\it Statistica Sinica}, {\bf5}, 727-736.

\item Zhu, L. P., Zhu, L. X. , Ferr\'{e}, L. and Wang, T. (2010). Sufficient dimension reduction through discretization-expectation estimation. {\it Biometrika}, {\bf 97}, 295-304.


\item Zhu, L. X. (2003). Model checking of dimension-reduction type for regression. {\it Statistica Sinica}, {\bf 13}, 283-296.

\item Zhu, L. X., Fujikoshi, Y. and Naito, K. (2001). Heteroscedasticity test for regression models.
{\it Sci. China Ser. A} {\bf 44}, 1237-1252.

\item Zhu, X. H., Guo, X., Lin, L. and Zhu, L. X. (2015). Heteroscedasticity Checks for Single Index Models. {\it Journal of Multivariate Analysis}, {\bf 136}, 41-55.


\end{description}

\
\newpage

\begin{table}[htb!]\caption{Empirical sizes and powers of $T^{DEE}_{n}$, $T^{ZH}_{n}$ and $T^{ZFN}_{n}$ for Example~1. \label{table1}
\vspace{0.75cm}}
\centering
 {\tiny\scriptsize\hspace{12.5cm}
\renewcommand{\arraystretch}{1}\tabcolsep 0.2cm
\begin{tabular}{cccccccccccccc}
\hline
&&\multicolumn{3}{c}{$T^{DEE}_{n}$}
& \multicolumn{3}{c}{$T^{ZH}_{n}$}
& \multicolumn{3}{c}{$T^{ZFN}_{n}$}\\
\hline
&$a/n$ &50 & 200& 400 &50 & 200& 400 &50 &200& 400\\
\hline
p=2
&0  &0.0370&0.0545&0.0525&0.0365&0.0415&0.0445&0.0455&0.0450&0.0550\\
&0.2&0.0535&0.1540&0.2680&0.0370&0.0490&0.1130&0.0620&0.0645&0.0865\\
&0.4&0.0900&0.3975&0.7675&0.0430&0.1610&0.5125&0.0710&0.1010&0.1790\\
&0.6&0.1530&0.6600&0.9870&0.0515&0.3595&0.8645&0.0720&0.1320&0.3815\\
&0.8&0.1975&0.8665&0.9960&0.0640&0.5605&0.9790&0.0780&0.1785&0.5385\\
&1.0&0.2520&0.9790&1.0000&0.0665&0.7260&0.9970&0.0820&0.1865&0.6780\\
 \hline
 p=4
&0  &0.0450&0.0535&0.0525&0.0105&0.0320&0.0700&0.0670&0.0650&0.0590\\
&0.2&0.0540&0.1570&0.3750&0.0377&0.0670&0.1920&0.0795&0.0750&0.1210\\
&0.4&0.0935&0.4990&0.8835&0.0305&0.0865&0.3525&0.0800&0.1160&0.1670\\
&0.6&0.1420&0.7895&0.9930&0.0265&0.3105&0.6810&0.0840&0.1230&0.2390\\
&0.8&0.2110&0.9265&1.0000&0.0365&0.4500&0.8050&0.1030&0.1970&0.3800\\
&1.0&0.2580&0.9650&1.0000&0.0565&0.5545&0.8535&0.1050&0.2900&0.4910\\
 \hline
 p=8
&0  &0.0460&0.0425&0.0535&0.0145&0.0225&0.0345&0.0420&0.0640&0.0580\\
&0.2&0.0620&0.1790&0.4380&0.0110&0.0550&0.1675&0.0500&0.0700&0.0975\\
&0.4&0.1270&0.4220&0.9530&0.0115&0.1175&0.3105&0.0580&0.0930&0.2040\\
&0.6&0.1740&0.8840&0.9870&0.0230&0.1910&0.4245&0.0555&0.1250&0.2605\\
&0.8&0.2090&0.9630&1.0000&0.0260&0.2490&0.5225&0.0660&0.1485&0.3575\\
&1.0&0.2390&0.9850&1.0000&0.0270&0.3640&0.6580&0.0715&0.1515&0.4080\\
 \hline
\end{tabular}
}
\end{table}

\begin{table}[htb!]\caption{Empirical sizes and powers of $T^{DEE}_{n}$, $T^{ZH}_{n}$ and $T^{ZFN}_{n}$ for Example~2. \label{table2}
\vspace{0.75cm}}
\centering
 {\tiny\scriptsize\hspace{12.5cm}
\renewcommand{\arraystretch}{1}\tabcolsep 0.2cm
\begin{tabular}{ccccccccccc}
\hline
&&\multicolumn{3}{c}{$T^{DEE}_{n}$}
& \multicolumn{3}{c}{$T^{ZH}_{n}$}
& \multicolumn{3}{c}{$T^{ZFN}_{n}$}\\
\hline
&$a/n$ &50 & 200& 400 &50 & 200& 400 &50 & 200& 400\\
\hline
\hline
$X\sim N(0,\Sigma_1)$
&0  &0.0580&0.0540&0.0535&0.0210&0.0565&0.0590&0.0730&0.0780&0.0900\\
&0.2&0.1115&0.5145&0.8985&0.0225&0.2295&0.5028&0.0850&0.1530&0.3415\\
&0.4&0.1635&0.8190&0.9775&0.0238&0.3045&0.5440&0.1055&0.2220&0.5540\\
&0.6&0.2120&0.8515&0.9835&0.0535&0.4047&0.5875&0.1215&0.2505&0.6310\\
&0.8&0.2615&0.8750&0.9830&0.1080&0.5005&0.6865&0.1300&0.2620&0.6685\\
&1.0&0.2820&0.9095&0.9930&0.1340&0.5925&0.7785&0.1360&0.2980&0.6740\\
 \hline
$X\sim N(0,\Sigma_2)$
&0  &0.0560&0.0450&0.0465&0.0375&0.0425&0.0605&0.0680&0.0590&0.0560\\
&0.2&0.1135&0.5655&0.9245&0.0520&0.1205&0.1800&0.0830&0.1855&0.4090\\
&0.4&0.1970&0.8575&0.9725&0.0980&0.3530&0.5300&0.0935&0.2955&0.5450\\
&0.6&0.2505&0.9060&0.9860&0.1085&0.5325&0.7165&0.1010&0.3000&0.6425\\
&0.8&0.2790&0.9205&0.9900&0.1390&0.6185&0.7875&0.1335&0.3120&0.6590\\
&1.0&0.3155&0.9335&0.9940&0.1615&0.6750&0.8105&0.1220&0.3310&0.6645\\
 \hline
\end{tabular}
}
\end{table}

\begin{table}[htb!]\caption{Empirical sizes and powers of $T^{DEE}_{n}$ for  Example~3. \label{table3}
\vspace{0.75cm}}
\centering
 {\tiny\scriptsize\hspace{12.5cm}
\renewcommand{\arraystretch}{1}\tabcolsep 0.2cm
\begin{tabular}{ccccc}
\hline
&&\multicolumn{3}{c}{$T^{DEE}_{n}$}
\\
\hline
&$a/n$ &50 & 200& 400\\
\hline
\hline
$X \sim N(0,\Sigma_1)$
&0  &0.0490&0.0530&0.0465\\
&0.2&0.0550&0.3280&0.6490\\
&0.4&0.1035&0.5965&0.9240\\
&0.6&0.1360&0.7250&0.9780\\
&0.8&0.1570&0.8110&0.9895\\
&1.0&0.1930&0.8665&0.9955\\
 \hline
$X\sim N(0,\Sigma_2)$
&0  &0.0615&0.0460&0.0480\\
&0.2&0.0615&0.2815&0.6445\\
&0.4&0.0965&0.5920&0.9300\\
&0.6&0.1395&0.7750&0.9910\\
&0.8&0.1475&0.8425&0.9920\\
&1.0&0.1830&0.9015&0.9980\\
 \hline
\end{tabular}
}
\end{table}

%

\end{document}